\documentclass[useAMS,usenatbib]{mn2e}
\usepackage{aasabbrev}
\usepackage{graphicx}
\usepackage{subfigure}
\usepackage{nicefrac}
\usepackage{color} 

\def\simpropto{\lower.2ex\hbox{$\; \buildrel \propto \over \sim \;$}}
\def\ltsim{\lower.5ex\hbox{$\; \buildrel < \over \sim \;$}}
\def\gtsim{\lower.5ex\hbox{$\; \buildrel > \over \sim \;$}}

\newcommand{\Msun}{{\rm M}_\odot}
\newcommand{\kms}{\, {\rm km\, s}^{-1}}

\title[Galaxy size trends as a consequence of cosmology]{Galaxy size trends as a consequence of cosmology}
\author[M.J. Stringer et al.]
{\parbox{\textwidth}{M.~J.~Stringer$^1$\thanks{martin.stringer@obspm.fr},  F.~Shankar$^{2,3}$, G.~S.~Novak$^1$,  M.~Huertas-Company$^{2}$,\\ F.~Combes$^1$ and B.~P.~Moster$^4$}\vspace{0.5cm}\\
\parbox\textwidth{$^{1}$Observatoire de Paris (LERMA), CNRS, 61, Av de l'Observatoire, Paris 75014, France\\
$^2$Observatoire de Paris (GEPI), CNRS, \& Universit\'{e} Paris Diderot, 4 Rue Thomas Mann, Paris 75013, France\\
$^3$Department of Physics and Astronomy, University of Southampton, Highfield, SO17 1BJ\\
$^4$Max-Planck Institut f\"{u}r Astrophysik, Karl-Schwarzschild Stra{\ss}e I, D-85748 Garching, Germany}}
\begin{document} 

\date{}

\pagerange{\pageref{firstpage}--\pageref{lastpage}} \pubyear{2014}

\maketitle

\label{firstpage}

\begin{abstract}
We show that recently documented trends in galaxy sizes with mass and redshift can be understood in terms of the influence of underlying cosmic evolution; a holistic view which is complimentary to interpretations involving the accumulation of discreet evolutionary processes acting on individual objects. Using standard cosmology theory, supported with results from the Millennium simulations, we derive expected size trends for collapsed cosmic structures, emphasising the important distinction between these trends and the assembly paths of individual regions. We then argue that the observed variation in the stellar mass content of these structures can be understood to first order in terms of natural limitations of cooling and feedback. But whilst these relative masses vary by orders of magnitude, galaxy and host radii have been found to correlate linearly. We explain how these two aspects will lead to galaxy sizes that closely follow observed trends and their evolution, comparing directly with the COSMOS and SDSS surveys. Thus we conclude that the observed minimum radius for galaxies, the evolving trend in size as a function of mass for intermediate systems, and the observed increase in the sizes of massive galaxies, may all be considered an emergent consequence of the cosmic expansion.
\end{abstract}
 

\section{Introduction}\label{Introduction}

Observational surveys of the radial extent of galaxies are now able to extend over many decades in stellar mass content \citep[e.g.][]{Ichikawa12,Bernardi13} and out to redshifts of 2 and above \citep[e.g.][]{Trujillo06,Ryan12,Barro13,Huertas13,vandeSande13}. These observations are allowing us to determine the relationship between stellar mass and radius, and follow the changes in this distribution across almost all of cosmic time. This has in turn prompted the question as to which physical processes could potentially cause the trends, and the changes in them with time. 	

Notably, there has been a great deal of assessment of the likely contribution to both from mergers between galaxies \citep[e.g.][]{Trujillo07}. Some calculations, using pair fractions \citep{Newman12} and cosmological predictions for merger rate \citep{Nipoti12} have tentatively concluded that such collisions cannot be the sole reason for the observed size evolution. Other estimates \citep{Bluck12,Lopez12} imply conversely that they are dominant. Other processes such as expansion after gas ejection have also been put forward to explain the evolution \citep[e.g.][]{Fan08}.

Meanwhile, there has been renewed interest in the relationship between the size and specific angular momentum of galaxies and that of their host structures. Classic ideas by \citet{Fall80} on the conservation of specific angular momentum from host structure to galaxy have been reenforced by  \citet{Kassin12}. Also, by matching the abundance of galaxies and the host structures predicted by theory, \citet{Kravtsov13} has shown that this implies a direct linear correlation between host and galactic radii. This is all consistent with theoretical galaxy formation pictures \citep{Mo98}, and refinements of this picture drawing on numerical simulations of galaxy formation have also been recently published \citep{Dekel13b}.

Motivated to connect these complementary research fields, the goal of this paper is to trace the effects of cosmic expansion through to the galaxy population; reviewing how the mean cosmic density is reflected in the density of collapsed cosmic structures, and understanding how this will  in turn be reflected by the densities -- and hence sizes -- of the central galaxies. Wherever possible, we will aim to follow this in terms of  accessible physical arguments. 

With this goal in mind, we begin by reviewing, in \S\ref{Theory}, the predictions from standard theory for the sizes of collapsed cosmic structures. To illustrate and support this, we then go on in \S\ref{Simulations} to study how these analytic arguments are borne out by the results of cold dark matter simulations of large cosmic volumes. 

In \S\ref{Galaxies}, we go on to consider the mapping from these host structures to central galaxies, beginning in \S\ref{rgal_rv} by reviewing the empirical and theoretical support for the proportionality, mentioned above, between host radius and {\em galactic} radius. Then, in \S\ref{MassContent}, we address the varying stellar mass content as a function of host mass, beginning again by reviewing analytic arguments which have been forwarded to explain this. In \S\ref{Mapping}, we then apply existing semi-empirical results \citep{Moster13} for this mass content to illustrate how the stellar mass correlation with host mass can equally be viewed as a correlation with host {\em radius} (using the theory reviewed in \S\ref{Theory}). 

This more holistic perspective is then bought to bear on some specific outstanding questions posed by the latest observational surveys. In particular, this demonstrates that similar mass galaxies at successive epochs are hosted by very different structures, leading to galaxy samples with very different radii - predicting an apparent evolution that is in line with observational measurements, shown in \S\ref{Application}. Finally, in \S\ref{Summary}, we summarise these results and the key theoretical arguments which support them.

\section{Sizes of collapsed cosmic structures}\label{Structures}

\subsection{The predictions of standard theory}\label{Theory}

In the standard theoretical picture of galaxy-scale structure formation, regions in the Universe which exceed a given critical overdensity will collapse to form final virialised regions with masses, $m_{\rm v}$ and radii, $r_{\rm v}$ determined uniquely by fundamental cosmological parameters and the time (or redshift, $z$) at which they ultimately collapse.
\begin{equation}
\frac{{\rm G}m_{\rm v}}{r_{\rm v}^3} \approx \frac{1}{2}\Delta_z\,H_z^2\label{Dv}
\end{equation}
Thus the family of structures which finally virialise at some particular redshift, $z$, carry densities which are an imprint of the universe of that epoch (i.e. $\propto H_z^2$) with higher order corrections to this dependence absorbed into $\Delta_{\rm z}$, the ratio of the final enclosed density\footnote{For spherically symmetric collapse, this has an early, matter-dominated value of $\Delta_{\rm z}\rightarrow18\pi^2$, but in $\Lambda$CDM this decreases to $\Delta_{\rm z}\approx 100$ for structures reaching virial equilibrium near $z\approx 0$.}  to the critical density.

This formalism from standard cosmology \citep[e.g.][and references therein]{White78,Cole91,Dutton11} immediately provides a simple, approximate prediction for the {\em instantaneous trend} that will exist in the population of structures extant at some given epoch in the universe. Namely, that  if the universe were populated by structures which have {\em just} virialised, we might expect to find the masses and radii of structures at any given epoch following a locus of constant density: 
\begin{equation}
R_v \approx  \left(\frac{2{\rm G}}{\Delta_z}\right)^{\nicefrac{1}{3}}\,\frac{M_{\rm v}^{\nicefrac{1}{3}}}{H_z^{\nicefrac{2}{3}}}~, \label{RvMv}
\end{equation}
where upper case symbols ($M_{\rm v}, M_\star$) refer to characteristic properties\footnote{The occupation function $M_\star(M_{\rm v})$ can be multi-valued whereas $m_\star$ and $m_{\rm v}$, referring to some individual, are single valued.} of a population or sample.

In reality, the structures will of course not all have instantaneously virialised together, but will have done so at a {\em range} of recent epochs, corresponding to a range of final densities; those with the lowest density being those which have only just collapsed, and structures with higher density having collapsed earlier and not yet been completely assimilated into any larger, less dense regions that have virialised around them. This will introduce a corelation which is {\em not quite} a constant density locus.

Variations in density, $\delta\rho/\rho$, are progressively less likely when considering larger and larger regions\footnote{i.e.$~\frac{{\rm d}\,\sigma_M}{{\rm d} M}<0~$ where $~\sigma_M^2\equiv \left\langle\left(\frac{\delta\rho}{\rho}\right)^2\right\rangle~$.}, but the amplitude of fluctuations grows with time\footnote{e.g. $\sigma_M(z) \propto (1+z)^{-1}$, in the matter-dominated era}. Because of this, a sample of higher-mass structures will have a later collapse time, on average, and thus carry a lower mean density than a sample of lower mass structures found at the same redshift. This means that structures at any given epoch would lie on a locus in the mass-radius plane that is both somewhat below the idealised constant-density locus (\ref{RvMv}), and also somewhat {\em steeper}.

These discussions benefit from characterising the radii of structures in the Universe at a certain time, represented by some given population, by the slope of the mass-radius correlation, $\beta$, the offset, $R_0$ at some given mass, $M_0$, and the evolution in the relation, $\gamma$.
\begin{equation}
R(M,z) \approx \frac{R_0}{{\left(1+z\right)}^\gamma}\left(\frac{M}{M_0}\right)^\beta\label{beta} 
\end{equation}
From this very brief discussion of the key elements of standard theory, we conclude that the population of structures in the mass-radius plane might be expected to have a slope close to, but a little higher than, $\beta\approx\nicefrac{1}{3}$, with an offset evolving with redshift as $\Delta_z^{-1/3}H_z^{-2/3}$, corresponding to $\gamma= 1$ at early times, dropping a little below this as $z\rightarrow 0$.

Is is useful to contrast the instantaneous trend, deduced above, with the trajectory of any particular {\em individual} region. These will not evolve {\em along} the near constant density loci described above, but move diagonally up {\em through} them; with each additional layer of accumulating matter around an existing structure virialising at a lower density.

\subsection{Comparison with simulations}\label{Simulations}

To illustrate the short theoretical review of \S\ref{Theory}, it is instructive to follow the sizes of collapsed structures in a cosmological simulation, looking at both the growth of individual regions and the trend that is found across the whole volume at any given snapshot. Such an illustration is shown in Fig.\,\ref{millennium}, taken from the publicly available results of the Millennium simulations \citep{Lemson06}.

In order to benefit from both the large volume of the original simulation \citep{Springel05} and the higher mass resolution of the Millennium II simulation \citep{Boylan09}, this figure combines results from both numerical experiments. Structures with $\log(m_v/\Msun)>13$ are taken from the MI volume of (500 Mpc/h)$^3$, which used a particle mass of $1.2\times10^9\Msun$. Lower mass structures are taken from the MII volume  ((100 Mpc/h)$^3$ and $9.4\times10^6\Msun$), and their number densities scaled to the larger volume. 

\subsubsection{Defining collapsed structures}\label{Definitions}

\begin{figure}
\includegraphics[trim=  21mm 100mm 159mm 20mm, clip, width=\columnwidth]{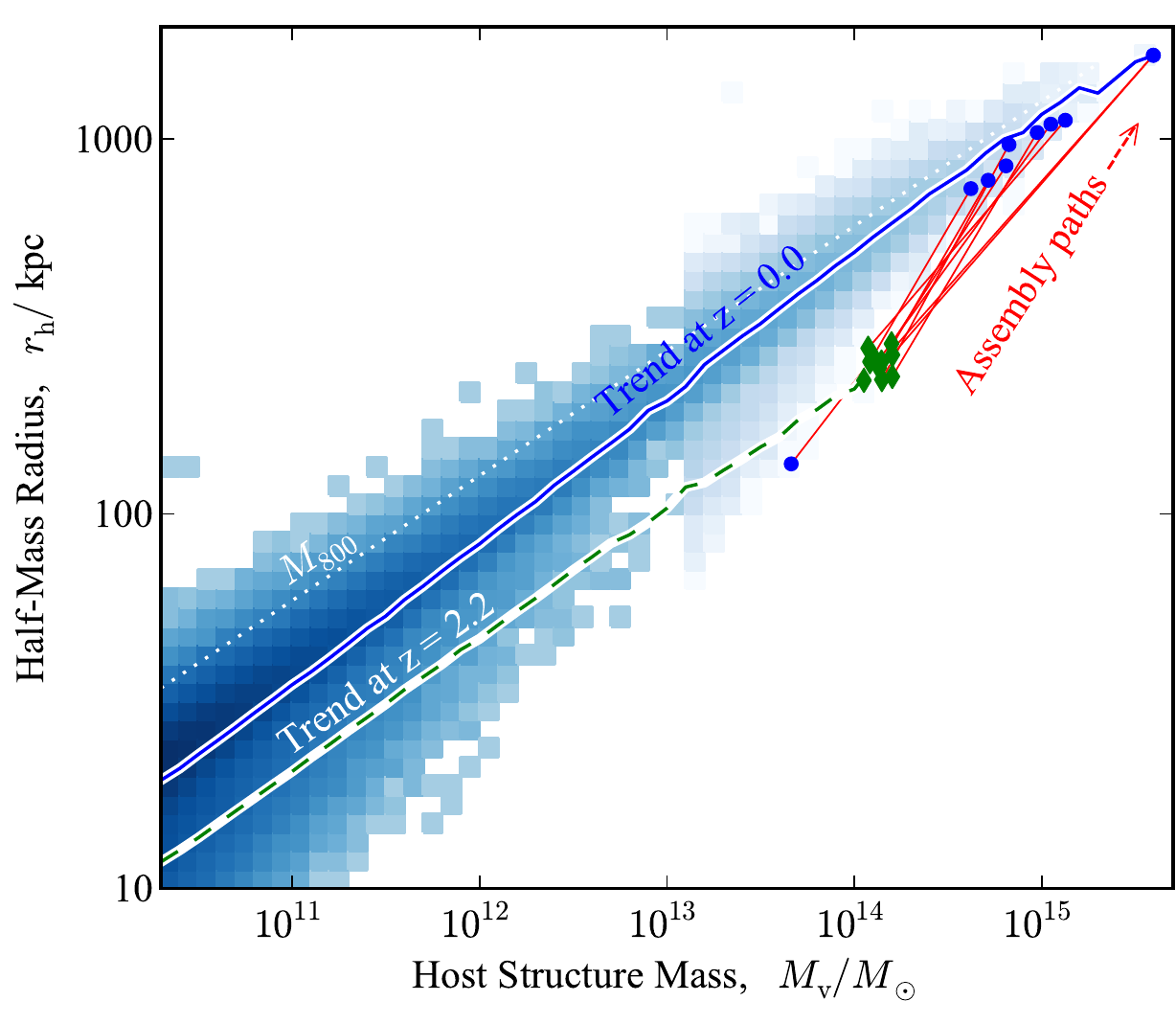}
\caption{An illustration of the trend in the population of structures seen at any given epoch and its evolution, contrasted with the evolution of individual regions. The blue shading indicates the virial masses and half-mass radii, $r_{\rm h}$ of all structures at $z=0$ in a (500Mpc/h)$^3$ simulation volume (see text). The solid line indicates the mean of log($R_{\rm h}$/kpc) for these structures as a function of stellar mass. The dotted line indicates a slope of constant mean density; 800 times critical for comparison with the half mass radius (as opposed to 200 which is usually chosen to represent the virial radius). The lower, dashed line shows the mean density of all structures in the same volume at $z=2.2$, with points showing the 10 most massive structures in the volume at this snapshot. These are linked to their $z=0$ descendants with thin solid lines. }\label{millennium}
\end{figure}

Before discussing the simulation results as constituting a confirmation of the basic theory in \S\ref{Theory}, it is worth a digression on how to actually define and measure the structures that collapse within it. The `host structure mass' plotted along the x-axis of Fig.\,\ref{millennium} was found in the standard way by associating all particles in an overdensity which lie within a distance $b/\bar{n}^{\nicefrac{1}{3}}$ of another particle, where $\bar{n}$ is the mean number density and $b$ is a free parameter, typically set to 0.2. This choice is traditionally motivated in order to enclose regions which contain an overdensity of $\Delta=200$. However, subtleties in this approach have recently been studied by \citet{More11}, who use percolation theory to show that it in fact selects regions with overdensities that vary depend on the individual density profile, and in practice tends to enclose regions with somewhat lower total overdensities (thus assigning structures with masses slightly larger than $m_{200}$). 

This is borne out by the structures in the high-mass (Millennium I) sample from Fig.\,\ref{millennium}. Where both the friends-of-friends mass, $m_{\rm FOF}$, and $m_{\rm 200}$ are available from the database, one finds $\langle m_{\rm FOF}/m_{\rm 200}\rangle=1.26$ and there are instances where the two masses differ by factors of ten or more. However the scatter in $\log(m_{\rm FOF}/m_{\rm 200})$ is small, of order 0.1. So, given that this is primarily just an illustrative analysis, we follow many previous authors in accepting the approximation $m_{\rm FOF}\approx m_{\rm 200}$.

Returning to the mass-radius corrleation, if we were to plot the radius enclosing this nominal overdensity (or a close proxy) then by their very definition all structures will fall {\em exactly} on a straight line with $\beta=\nicefrac{1}{3}$ (corresponding to eqn. \ref{Dv} with the constant value of $\Delta=200$). The locus will indeed move upwards according to (\ref{beta}), but only because we have defined it this way and the mean density in the simulation is falling (not {\em necessarily} because the structures are actually growing). So the mass-radius relation of structures would effectively be just a plot of our chosen definition of a structure, and thus cannot be used to corroborate independently this aspect of structure formation theory. How to avoid this enforcement of analytic theory on the interpretation of the experiment, and find a completely objective definition for a virialised region, is not obvious. 

But recent renewed interest in the density profiles of simulated structures by \citet{Ludlow13}, exploring the density averaged over varying fractions of the entire structure, helps to resolve this issue. They find that {\em central} density\footnote{Specifically, the density inside the radius at which ${\rm d}^2m/{\rm d}r^2=0$.} certainly correlates extremely strongly with the cosmological density at formation time\footnote{The definition of the formation time is still a little subjective as is when the `total' mass (which remains a function of free parameters) is equal the final central mass (as defined above).}, confirming the seminal work of \citet{Zhao03}. This lends some numerical assurance to the analytic discussion from \S\ref{Theory}, particularly as it is these central dynamics are those most relevant for the incumbent galaxy. So, in recognition of this, Fig.\,\ref{millennium} shows the half mass radii, $r_h$, of structures in the simulation, which represents the density of the simulated structures whilst also avoiding the trap of being entirely driven by our analytic expectations.

\subsubsection{The instantaneous trend in the population}

Given these slight caveats in \S\ref{Definitions}, the distribution of radii in Fig.\,\ref{millennium} does indeed appear to follow the predictions of basic theory. The mean of $\log (r_{\rm h}/{\rm kpc})$ as a function of stellar mass at $z=0$ is shown by the solid line, and does indeed have a slope a little steeper than $\beta=\nicefrac{1}{3}$; lower mass objects having, on average, collapsed earlier and thus reflecting the denser cosmic environment at this time.

The same mean is also shown for structures at $z=2.2$ (a dashed line in the figure) and this is indeed found to have a lower offset, as discussed above. The slopes are similar, but not exactly the same. This uncovers the additional interesting detail that the slope of the mass radius relation for structures steepens as the universe evolves. The reason for this is that, in an older universe, there can be a greater difference in characteristic collapse epoch from low to high mass. At earlier times, the structures are closer to lining up around a locus of constant density, and the trend deviates gradually from this as the universe continues to evolve.

\subsubsection{The growth of individual regions}\label{Individuals}

In addition to the distribution of the population as a whole, Fig.\,\ref{millennium} shows the virial masses and half-mass radii of the 10 most massive systems at $z = 2.2$, and links these to those of their descendants at  $z = 0$. One can see that these `paths' generally go upwards through the set of mean density trends at each epoch; the path of `individual' evolution is much steeper than the static trend, as argued in section \ref{Theory}.

As a caveat to this, it must be pointed out that the structures picked out at the two epochs  -- even in this simple CDM case -- are difficult to view as being `the same object', especially not when the time between them is such a large fraction of the Hubble time. In this example, all the structures at  $z=2.2$ represent less than half, and some not even a tenth of the mass of the eventual redshift zero descendant. It is tenuous, at best, to discuss them as being the same entity. Even just this simple statistic alone should persuade us that great care is needed when making any connections between descendants and progenitors in a hierarchical formation scenario, particularly across such vast expanses of time.

Further insights into this issue can be gained by simply following the rank order of the 10 highest-mass structures in Fig.\,\ref{millennium}. For example, one of these 10 becomes a satellite of another and {\em decreases} in mass. Another two are deemed to have merged entirely (becoming in fact the most massive halo at z=0). The remaining halos remain distinct, but none of them remain in the top 10 at $z\sim0$. Indeed, they are not even all in the top 100; the least massive of their descendants ranking only in the high 300s.

Though the fates of these 10 structures is just an anecdotal example, the simplicity of the case hopefully makes it very clear that associating low- and high-redshift structures by matching their rank order by mass is an assumption that is certainly not supported in detail at the level of structures in general. Whether or not the {\em galaxies} at the center of these collapsed regions might follow such an assumption more closely is a more difficult question to address. 

\section{From host structures to galaxies}\label{Galaxies}

In \S\ref{Theory} we reviewed the very basic expectations of mass--size correlation for collapsed structures in standard cosmology. To advance this discussion and convert this into an equivalent prediction for the galaxies within them, we can begin by briefly reviewing the relationship between host radius and galactic radius (\S\ref{rgal_rv}) and host mass and stellar mass (\S\ref{MassContent}). This will hopefully reveal to what extent the evolutionary behavior deduced for structures in general in \S\ref{Structures} is retained by the central objects,  to what extent it is broken, and to identify the key physical limit or process which drives each case.

\subsection{From host radii to galactic radii}\label{rgal_rv}

The discussion so far has been restricted to dark-matter dominated structures (\S\ref{Structures}) and the evolution of their {\em virial} radii, or proxies like the half mass radius (Fig.\,\ref{millennium}) or $R_{\rm 200}$. But for our discussion to be useful to understanding galaxy size variation, we need now to review our theoretical and/or empirical knowledge of how these galaxies trace the size of their host structures.

Accumulating empirical and theoretical hints suggest the existence of a correlation between galaxy size and virial radius. \citet{Kravtsov13} have compared the sizes of galaxies galaxies of all morphological types from a collection of observational samples \citep{Leroy08,Misgeld11,Zhang12}, with the virial radii of structures competing to dark matter haloes  in a simulated volume of $(250 {\rm Mpc}/h)^3$ \citep{Klypin11}. The comparison was made by matching their respective cumulative abundance per unit volume\footnote{That is, associating galactic radius $R_{\rm gal}$ with host radius $R_{\rm v}$ if: \\$n_{\rm gal}(>R_{\rm gal}) = n_{\rm hosts}(>R_{\rm v}$).} and was statistically consistent with the two radii being directly proportional: 
\begin{equation} 
R_{\rm gal}=\lambda R_{\rm v}, 
\end{equation}
with $\lambda\approx 0.015$, independent of galaxy morphology.  

As pointed out by the author, this result appears to strongly support the picture of \citet{Mo98} where the mean specific angular momentum, $j$, of material scales with the host structure: $j  \approx r_{\rm v}v_{\rm c}$. Thus if the material cools until supported by bulk motion, and the rotation curve out at these large radii is close to flat (or changes by a consistent factor in all structures), it will settle at $r_{\rm gal}\propto r_{\rm v}$. Further, complimentary reinforcement of this view has been published recently by \citet{Kassin12}, who show that the directly measured specific angular momentum of galactic systems, as a function of characteristic velocity (in the range  $125<v_{\rm c}<315$), matches the same trend for simulated dark-matter dominated structures.

In addition to the basic theoretical picture of conservation of specific angular momentum, more specific physical processes to actually transfer material from the outskirts of dark matter haloes to galaxy scales have also been proposed in the literature. For example, \citet{Dekel13b} have recently summarized the results of high-resolution hydro-cosmological simulations of massive galaxies at $z>1$. They confirm that baryons falling along cosmic filaments can penetrate down to the inner regions of the central proto-galaxy, feeding the continuous formation of a gas-rich, clumpy disc. In particular, they emphasize that the disc radius maintains a nearly constant proportionality of a few percent with its host virial radius during the full evolution of the simulations, a result in remarkably good agreement with the empirical findings discussed above.

On the other hand, it has also been recognized that most galaxies which dominate the high-mass end of the stellar mass function completed almost all their star formation a long time ago ($z\gtsim 1$). Thus their subsequent evolution could have only happened via a sequence of mergers with incoming satellites, a possibility which has been put forward to explain also their apparent strong size growth \citep[e.g.][]{Naab09}. It has also been extensively discussed in the present literature that mergers may not be entirely sufficient to explain the size growth at fixed stellar mass for massive spheroids \citep{Nipoti12, Huertas13, Shankar13}. Understanding how broad trends in the galaxy population arise, in this way, from cumulation of many discreet individual evolutionary events is a valuable and complex theoretical challenge, and has been taken up by many authors \citep[e.g.][]{Somerville08,vanderWel09,Hopkins09,Dutton11,Cassata11,Cassata13}.  

We do not confront this debate directly here. The contrasting, but complimentary perspective we wish to promote is that mergers and diffuse gas accretion alike are, ultimately, both transporting mass and specific angular momentum from the outer parts of a collapsed structure to its central galaxy. As such, both can be thought of as different modes by which galaxy growth tracks the mass and structural growth of its host dark matter halo. This would be consistent with the conclusions of \citet{Carollo13}, whose analysis of non-star-forming elliptical galaxies in the COSMOS sample concludes that average sizes roughly scale with the average density of the universe at the time when their star formation ceased, an idea that has also been supported by recent theoretical modelling \citep{Posti14}.

To further explore the implications of this more holistic view, we investigate in the following sections the effects of cosmic evolution on the galaxy population for the scenario where galaxies do indeed track their hosts, examining the consequences of the {\em canonical assumption} $r_{\rm gal}\propto r_{\rm v}$. As well as carrying significant observational and theoretical support reviewed above, this choice is additionally motivated by its simplicity, allowing direct cosmological effects on the galaxy population to be followed clearly throughout physical arguments and accessible calculations. This is therefore presented as a theoretical reference point, allowing the results of more complex models to be interpreted in terms of accessible analytic calculations which connect more palpably to the cosmology.

\subsection{From host mass to central galaxy}\label{MassContent}

In \S\ref{rgal_rv}, we have reviewed the evidence in support of a proportionality between galactic radii and the radii of the their host structures. Given that the evolution of the latter can be understood from cosmology (\S\ref{Theory}), it remains only to understand the correlation between host {\em mass} and galactic mass, in order to appreciate the cosmological effects on the galactic stellar mass--radius relationship, which is our goal. 

So in this section we will begin by briefly reviewing the theoretical understanding of the varying stellar mass content of structures, dividing the discussion into three regimes, corresponding to the three different physical effects understood to be broadly responsible in each case. In \S\ref{Mapping} we go on to use semi-empirical occupation functions to illustrate how this static correlation between $M_\star$ and $M_{\rm host}$ leads, via the cosmology, to the evolving correlation between stellar mass and radius which we set out to understand.

\subsubsection{The lowest mass galaxies}\label{LowMass}

In the limiting case of very low mass galaxies, fractional stellar mass content is extremely small; observational estimates implying barely thousandths of the total mass in some cases \citep[e.g.][]{Walker09}. Physically, this can be understood as due to inefficient cooling at these low virial temperatures, exacerbated by the potency of supernovae in these smaller potential wells.

The inefficiency is linked to the cut-off in atomic  cooling at temperatures below $10^4$K, enforced by the background radiation from the first stars, thought to suppress cooling at similar temperatures \citep[e.g.][]{Okomoto08}. Though the cut-off in the atomic cooling rate, in particular, is very sharp, it does not emerge as a clear cut off in velocity dispersion for the galaxy population. 

As the eventual structure assembles in an evolving cosmological environment, its virial temperature will fluctuate. 	For example, a structure with a final virial temperature that exceeds this threshold may have recently formed from progenitors which were all below it, and thus have only just begun to form efficiently a central galaxy. Conversely, a structure which would be deemed to be below this threshold for galaxy formation, based on it's final velocity dispersion, may well have had efficiently-cooling progenitors in the past and thus contain a significant galaxy \citep[see][for details]{Stringer10}.

\begin{figure}
\includegraphics[trim= 4mm 136mm 4mm 4mm, clip, width=\columnwidth]{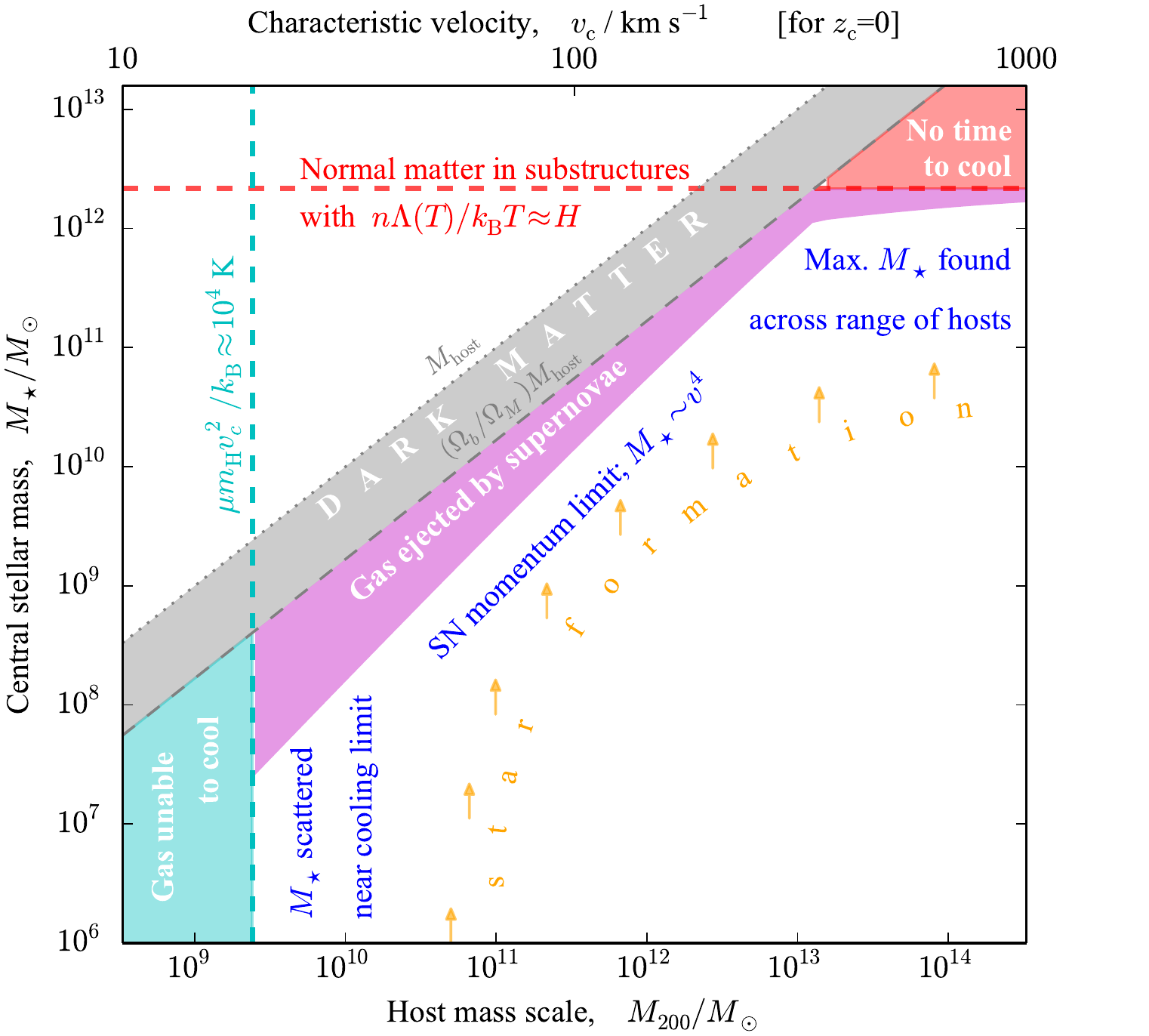}
\caption{A basic illustration of some of the theoretical constraints on the central stellar mass content of cosmic structures.  Shaded regions indicate the material which is prevented from coalescing from the outer regions of the structure onto the central galaxy. Star formation can then only drive the galaxy population to the domain just below these limits, creating a correlation that approximates into three regimes, corresponding to three basic physical limits.  Each of these is explained in more detail in the respective sections \ref{LowMass}-3, but we reenforce here that this is based on recently collapsed structures ($z_c \approx 0$, as indicated on the top axis). Hierarchical assembly creates some deviation from this, as explained in the text, and more importantly the entire relationship will evolve somewhat from one redshift to another.}\label{sketch}
\end{figure}

Thus, having arrived there by virtue of a variety of assembly histories, structures with virial temperatures on and around the threshold will be populated by galaxies whose stellar masses correlate very weakly, if at all, to their hosts'. So in the familiar stellar mass--velocity dispersion plane, illustrated for reference here in Fig. \ref{sketch}, the low mass limit of the population will be widely scattered in stellar mass, around a minimum characteristic velocity scale.

Because the virial radius and characteristic velocities are related directly by the cosmology (as with the mass, eq. \ref{Dv}), the characteristic velocity, $v_{\rm c}$ will correspond also to a size scale, given by:
\begin{equation}\label{rv}
r_v = \frac{v_{\rm c}}{\left(\nicefrac{1}{2}\Delta_z\right)^{\nicefrac{1}{2}}H_z}~.
\end{equation}
Thus the limiting virial temperature for galaxy formation  will have a corresponding minimum virial radius (varying with redshift), and structures with radii around this value can contain a wide range of values of stellar mass, as discussed.

Turning then to the stellar mass-radius relation, these arguments suggest that this should be almost flat; structures with widely ranging stellar mass content all belonging to structures with approximately the same common virial radius. Following (\ref{rv}), and noting that empirical \citep[e.g.][]{Walker09} and theoretical\footnote{e.g. structures with a threshold virial temperature of $10^4{\rm K}$ have associated velocity $(3{\rm k}_{\rm B}\times10^4{\rm K}/\mu m_{\rm H})^{\nicefrac{1}{2}}\approx20\kms~.$} indications support $v_{\rm min}\sim 10\kms$, this common mimumum radius at which the stellar mass-virial radius relation flattens would be expected to be roughly:
\begin{equation}
R_{\rm 200, min}(z) \sim\frac{30\,{\rm kpc}}{\left(1+z_{\rm c}\right)^{3/2}}~,\label{rmin}
\end{equation}
where $z_{\rm c}$ refers to the collapse redshift of the structure (not the redshift at which the galaxy is observed). This limit appears at the low-mass extreme in Fig. \ref{sketch_radius}, which shows how the natural limits on stellar mass content (Fig \ref{sketch}) translate to limits on host radius.

These simple arguments, leading to (\ref{rmin}), imply the clustering of systems with varying magnitude (or stellar mass) around a common minimum {\em host} structure radius. Following the review of \S\ref{rgal_rv} we can also equate this to an equivalent galactic radius, on the basis that it is the residual specific angular momentum from the host structure, no matter how redistributed or disoriented, that is ultimately responsible for retaining the physical extent of the central galaxy. 

If we apply the collective behaviour $R_{\rm gal} \approx \lambda R_{\rm v}$ to the minimum structure radius derived in (\ref{rmin}) we find that the theoretical expectation for the limiting physical scale for galaxies (forming at, and around, virial temperatures of $T_{\rm c}\approx 10^4$K) is of order:
\begin{equation}
R_{\rm gal, min} \sim \frac{\lambda}{H_z} \left(\frac{6{\rm k}_{\rm B}T_{\rm c}}{\mu m_{\rm H}\Delta_z}\right)^{\nicefrac{1}{2}} ~~\sim \frac{440 {\rm pc}}{(1+z_{\rm c})^{\nicefrac{3}{2}}}~,\label{rgal_min}
\end{equation}
where we have substituted, by way of example, the value of $\lambda=0.015$ found by \citet{Kravtsov13}. This theoretical expectation can be compared with local observations of satellite galaxies, where indications of a limiting radius have indeed been presented, first by \cite{Belokurov07} who showed galaxies ranging over 8 magnitudes in the V-band all occupying a lower limit in half light radius at around 100--300pc.

To interpret this in the context of the basic theory, it is important to remember that structures virialise at {\em recent} (not necessarily current) epochs. The effect is all the more important for these lower mass systems systems because many are satellites. Once accreted into a larger virialised region, satellites become de-coupled from the cosmology, so their overdensity would be relative to the cosmic value at their {\em accretion} time, rather than the current time. 

This is illustrated in  Fig. \ref{sketch_data} which allows us to compare the observed radii of the local dwarf spheroidals \citep{Walker09} to the 1st-order theoretical estimate (\ref{rgal_min}), indicating that a collapse epoch, $z_{\rm c}\sim 0-3$, is entirely consistent with the \citet{Belokurov07} result. If this explanation is correct, such observations of families of satellites in general could in future be considered a rough estimate of their group's principle formation epoch. 

\subsubsection{Intermediate mass galaxies}

At intermediate masses, the relationship between galaxy and host structure tightens greatly. Notably, for disks, there is the well-established correlation between stellar mass and characteristic velocity very close to $M_\star\propto v_{\rm max}^4$ over two decades in stellar mass \citep[e.g.][]{Miller13}. Physically, this relationship can be understood in terms of a momentum and energy budget from supernovae and stellar wind-driven outflow that is similar for all systems, but a gravitational potential barrier to outflow which varies greatly across the range of structure masses in which galaxies are found \citep{Mathews71,Larson74}.

A simple analytic estimate of the combination of these physical effects ought ideally to take into account a range of outflow velocities \citep{Stringer12} and also varying gas surface density, gas fraction and disk height \citep[e.g.][]{Creasey13}. However, the basic argument is that {\em the same kind of supernovae are acting in very different potential wells}. This is surely correct at some level, and can be quantified in a basic, but instructive, way by arguing as follows: 

If some mass, $M_{\rm out}$, has successfully escaped from the region of the galaxy then, to have done so, it must at some earlier stage been moving out with mean velocity $\sim v_{\rm c}$. In the approximation that these early stages tend to carry a fixed specific outward momentum budget per mass of stars formed, $v_{\rm w}$, we can then write\footnote{This is approximate, but a differential version of the argument, $M_{\rm out} \approx \int\dot{M}_\star(t) v_{\rm w}/v_{\rm c}(t){\rm d}t$, including hierarchical formation, can be seen to lead to the same basic scaling \citep{Stringer10}.}  $M_{\rm out} v_{\rm c}\approx M_\star v_{\rm w}$. Finally, in the regime of interest in this section (intermediate host mass with very effective cooling), we also have the constraint\footnote{More formally, this should read $f_\star^{-1}M_\star + M_{\rm out} = (\Omega_{\rm b}/\Omega_{\rm M})M_{\rm v}$, where $f_\star\equiv M_\star/M_{\rm gal}$. For simplicity, here, we follow the argument through with $f_\star\sim 1$. } from cosmology: $M_\star + M_{\rm out} = (\Omega_{\rm b}/\Omega_{\rm M})M_{\rm v}$. Together these imply that even in the hypothetical limit where all material cools at some stage onto the galaxy, structures are subject to a limiting central stellar mass:
\begin{equation}\label{TF}
M_\star \approx\frac{\Omega_{\rm b}/\Omega_{\rm M}}{ \left(\nicefrac{1}{2}\Delta_z\right)^{\nicefrac{1}{2}}{\rm G}H_z}~~\frac{ v_{\rm c}^4}{v_{\rm w} + v_{\rm c}}~.
\end{equation}
In this intermediate regime where cooling onto the central galaxy is expected to be very effective, the stellar mass--characteristic velocity relation might be expected to approach this limit, $M_\star\propto v_{\rm c}^4$, at low to intermediate masses, moving towards $M_\star\propto v_{\rm c}^3$ as circular velocities approach the specific momentum budget from supernovae (at which point cooling limitations also begin to apply again, as discussed in the next section, \S\ref{Massive}).

\begin{figure}
\includegraphics[trim= 4mm 3mm 4mm 143mm, clip, width=\columnwidth]{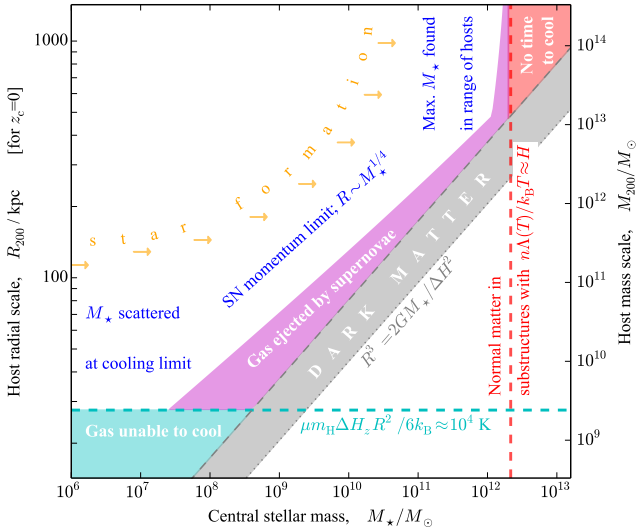}
\caption{A demonstration of how the basic physical limits which constrain the familiar relationship between stellar mass and host structure mass, shown in Fig.\ref{sketch}, also create a well-defined relationship between stellar mass and host {\em radius}. Here, it is even more important to emphasise that the entire correlation translates upwards as the universe, and the structures, evolve (changing $z_{\rm c}$). This evolution is derived analyticaly in \S\ref{MassContent} and its emergence in the context of a scattered population investigated in \S\ref{Mapping}.}\label{sketch_radius}
\end{figure}

In the standard assumption that the characteristic velocity of the galactic system is closely matched to that of the host, we can use this feedback-driven relationship in {\em velocity} (\ref{TF}) to write the resulting correlation between stellar mass and host {\em radius}, for galaxies at the lower end of this mass range:
\begin{equation}
GM_\star \approx \frac{\Omega_b}{\Omega_M}\left(\frac{\Delta_z}{2}\right)^{\nicefrac{3}{2}} \frac{H_z^3R_{\rm v}^4}{v_{\rm w}}\hspace{0.5cm}{\rm or}
\hspace{0.5cm}
R_{\rm v}\propto\frac{M_\star^{\nicefrac{1}{4}}}{H_z^{\nicefrac{3}{4}}}~.\label{rv_int}
\end{equation}
At higher stellar mass, there will eventually be a further transition to the most massive regime, which we will discuss further in \S\ref{Massive}, where cooling limits become important again. At this transition, the $M_\star-M_{\rm v}$ correlation flattens, as can be seen by reference to Fig. \ref{sketch} (which shows the example case $v_{\rm w}\approx 300\kms$).

This flattening in stellar mass content translates to a {\em steepening} of the mass--radius correlation, illustrated in Fig. \ref{sketch_radius}. So there will be some range in mass, and radius, in which galaxies track the evolution of host structures themselves, recovering $\beta\approx\nicefrac{1}{3}$ and mirroring eqn. \ref{Dv}:
\begin{equation}
R_{\rm gal} \propto\frac{{M_\star}^{\nicefrac{1}{3}}}{H_z^{\nicefrac{2}{3}}}~. \label{rgal_int}
\end{equation}
In summary of \S\ref{MassContent} so far, the trend in virial radius as a function of stellar mass might be expected to rise from the flat relation, $\beta\sim 0$, argued for in \S\ref{LowMass}, to $\beta\approx\nicefrac{1}{4}-\nicefrac{1}{3}$ as argued in this section. This prediction for the host radii runs alongside observations of {\em galactic} radii. \citet{Ichikawa12}, for example, find $\beta\approx 0.1$ for all galaxies in the $7<\log(M_\star/\Msun)<10$, and simple regression fits to the SDSS sample of \citet{Bernardi13} yield $\beta=0.21$ for $9<\log(M_\star/\Msun)<10$, steepening to $\beta= 0.29$ for $10<\log(M_\star/\Msun)<11$. This slope then rises rapidly for the most massive galaxies, as will be discussed in \S\ref{Massive}.

\begin{figure}
\includegraphics[trim= 3mm 3mm 4mm 4mm, clip, width=\columnwidth]{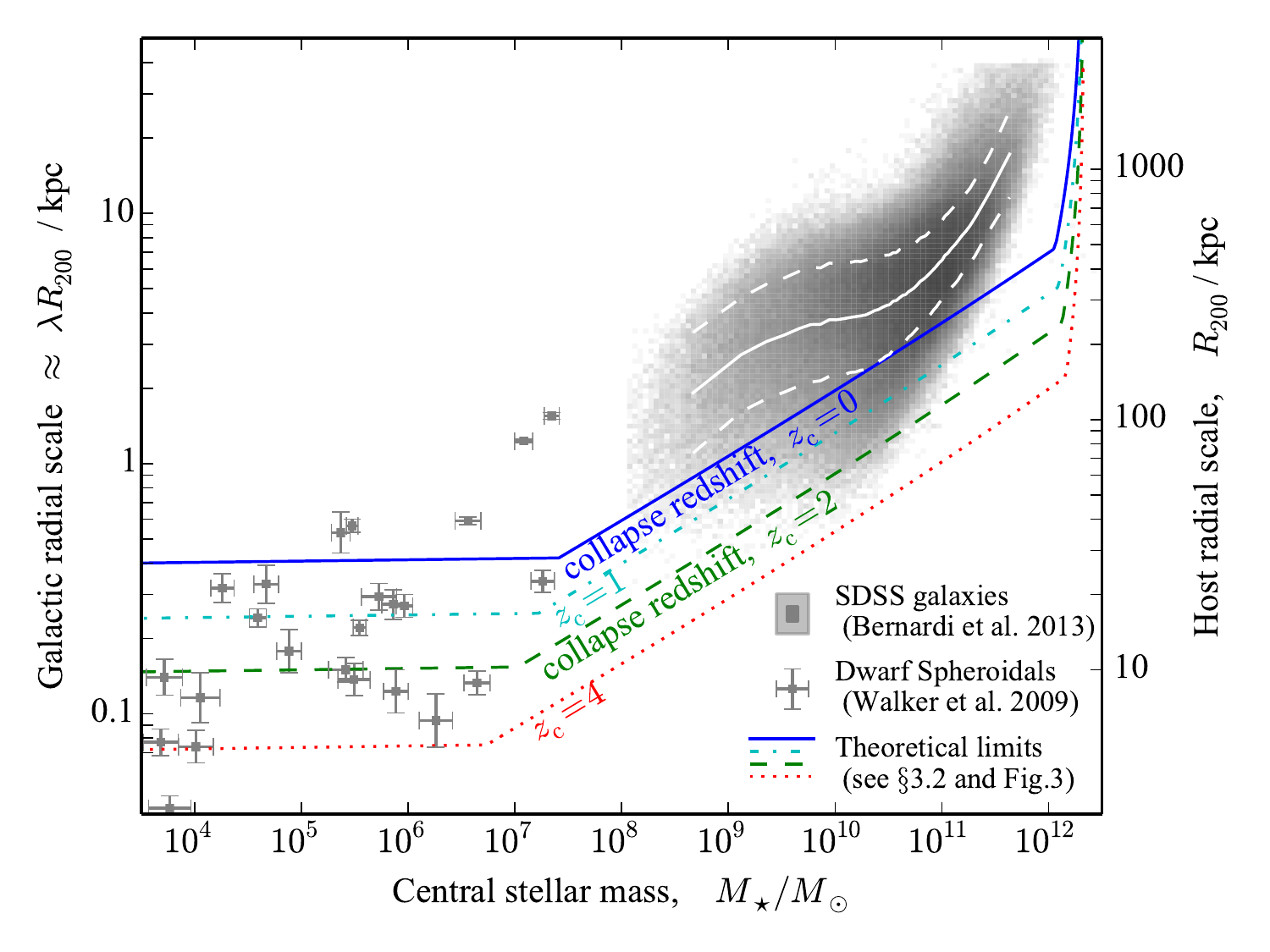}
\caption{A comparison between the basic theoretical limits sketched in Fig. \ref{sketch_radius} (see \S\ref{MassContent}) and observations of galactic stellar masses and half light radii from local dwarf spheroidals \citep[][errorbars]{Walker09} and the SDSS \citep[][shading]{Bernardi13}. Lines within the shaded region indicate the mean and standard deviation of $\log(r_{\rm gal}/{\rm kpc})$ within each of 30 stellar mass bins of $\sim 8000$ galaxies. The four separate lines correspond to theoretical limits for four different collapse redshifts, as labelled, corresponding to $R\sim M_\star^{\nicefrac{1}{4}}$ (\ref{rv_int}) in the intermediate mass range.}\label{sketch_data}
\end{figure}

For a visual comparison, the locus of the SDSS galaxies are included in Fig. \ref{sketch_data}. The comparison appears broadly consistent with these theoretical limits {\em in the context of the hierarchical picture}. The observations line up diagonally through the loci of constant collapse redshift, corresponding to lower mass structures which, on average, collapse earlier than those at higher mass. This can be interpreted as the galactic analogue of the structural trend that emerges from hierarchical formation, discussed in \S\ref{Introduction} and shown in Fig. \ref{millennium} to exist for simulated structures.

\subsubsection{The most massive galaxies}\label{Massive}

At the highest end of the stellar mass range, the $M_\star-M_{\rm v}$ relation quoted above continues to flatten as the correlation between the host mass (or circular velocity) and the stellar mass of the central galaxy begins to no longer hold, and is eventually lost. Structures exist locally which are deduced to contain $10^{15}\Msun$ and higher \citep[e.g.][]{Dai12,Lidman12}, but though these may be thousands of times the mass of structures like that which hosts our own galaxy, for example, the central galaxies are nowhere near this many times more massive. Only small differences in stellar mass are found across a wide range of host structure masses at the most massive end of the population.

The physical reason for this is that though these larger structures do host more gas, there simply has not been enough time for it to radiate its energy and coalesce into the centre of the region  \citep{Rees77}, a limit which becomes relevant for structures as their cooling time approaches, and eventually exceeds, the Hubble time:
\begin{equation}
\frac{k_{\rm B}T_{\rm v}}{n_{\rm v}\Lambda(T_{\rm v})} \sim \frac{1}{H}\label{T_max}~,
\end{equation}
where $n_{\rm v}$ is the number density of normal matter enclosed by the structure\footnote{A rather better expression of this condition might distinguish time at collapse, $t_{\rm c}$ and time of observation, $t_{\rm o}$:
\begin{equation}
\frac{k_{\rm B}T_{\rm v}(t_{\rm c})}{n_v(t_{\rm c})\Lambda(T_{\rm v}(t_{\rm c}))} \gtsim t_{\rm o}-t_{\rm c}~,\label{T_max_c}
\end{equation}
whereas the simplified version (\ref{T_max}) considers the case where {\em even} an entire Hubble time would be insufficient for the structure to cool. This generosity is countered somewhat by the approximation on the l.h.s., $T(t_{\rm c})\sim T(t_{\rm o})$, which is also an overestimate (the l.h.s $\propto T^{\nicefrac{1}{2}}$ and for a growing structure we would expect $T(t_{\rm c})<T(t_{\rm o})$). So for the present purpose of lending basic quantitative support for the approximate limiting scale, (\ref{T_max}) will suffice.
}.
This is of course a very basic argument. But as most considerations that are being neglected (notably, of course, further heating of this gas by radiation from AGN) will act to further reduce it, it remains quite robust as an upper limit on regions which can collapse to produce a single, dominant central galaxy.

Once structures approach, and exceed, this approximate limiting mass and temperature scale (\ref{T_max}), additional accreted mass will no longer be reflected in the central stellar content. The condition therefore creates in turn an effective physical limit on maximum stellar mass of any one galaxy (irrespective of what superstructure it may be imbedded in). This constraint appears as the dashed horizontal line in Fig. \ref{sketch}.  These most massive galaxies that can be produced in nature are therefore to be found at the centre of a {\em variety} of host structures, which all lie above an approximate cut-off structure mass scale. So at this highest-mass end the well-defined correlation between host structure and central galaxy begins to break down.

To determine the expected mass-radius relation that results from this, it is more suitable to discard the discussion of a well-defined slope  (tending to $\beta\rightarrow\infty$, as illustrated in Fig. \ref{sketch_radius}), and consider the population of galaxies to be {\em scattered} around the maximum natural limit for stellar structures, and realise that they are hosted by structures with a much wider distribution in host mass. This will have crucial implications for the host radii of galaxies at this same maximum mass, but difference epochs. 

For example, if we are interested in understanding the mean host properties of all galaxies in a given stellar mass range, we are sampling all structures which host such galaxies. For intermediate galaxies, where host and stellar mass are more tightly correlated, this sample of host structures may all be quite similar ($\delta M_{\rm v}/\delta M_{\star} \sim 1$). But, following the arguments above, the range of host structures sampled by the highest stellar mass bin will be much greater ($\delta M_{\rm v}/\delta M_{\star} >> 1$). If we wish to seek an analogue of equations (\ref{rgal_min}) and (\ref{rgal_int}), but for high mass galaxies, we cannot simply ask the question: 
\begin{quotation}
``what is a {\em typical} host radius for galaxies of this stellar mass?"
\end{quotation} 
but should ask: 
\begin{quotation}
``what is the {\em mean radius of all structures} which could host galaxies of this stellar mass?"
\end{quotation}  
Mathematically, then, we seek something of the form:
\begin{equation}
\left< R_{\rm gal}(M_\star)\right> \approx \frac{\lambda}{N(>M_{\rm lim})}\int_{M_{\rm lim}(M_\star)}^\infty\frac{{\rm d}N}{{\rm d}M_{\rm v}}\, R_{\rm v}\,{\rm d}M_{\rm v} \label{Rmax}
\end{equation}
To address this question analytically, with an eye to comparison with observations, one would need to take account of the varying survey volume with redshift, solve a version of (\ref{T_max_c}) which correctly incorporates hierarchical formation, and also take into account the contribution to stellar mass content from substructures.

To retain the simplicity of \S\ref{LowMass} \& 2, we therefore leave this more detailed analysis for future work, and rest with a basic theoretical argument in support of the (observed) limiting stellar mass. For the present purposes, of pursuing the consequences of a basic fundamental limit on stellar mass for galactic size evolution, we can confront these questions more quantitatively with the aid of an illustrative mock sample in \S\ref{Mapping}.

\subsection{Illustration using abundance matching results}\label{Mapping}

\begin{figure*}
\subfigure[The inferred mapping between stellar mass and host mass.]{\includegraphics[trim= 34mm 0mm 212mm 18mm, clip, width=0.49\textwidth]{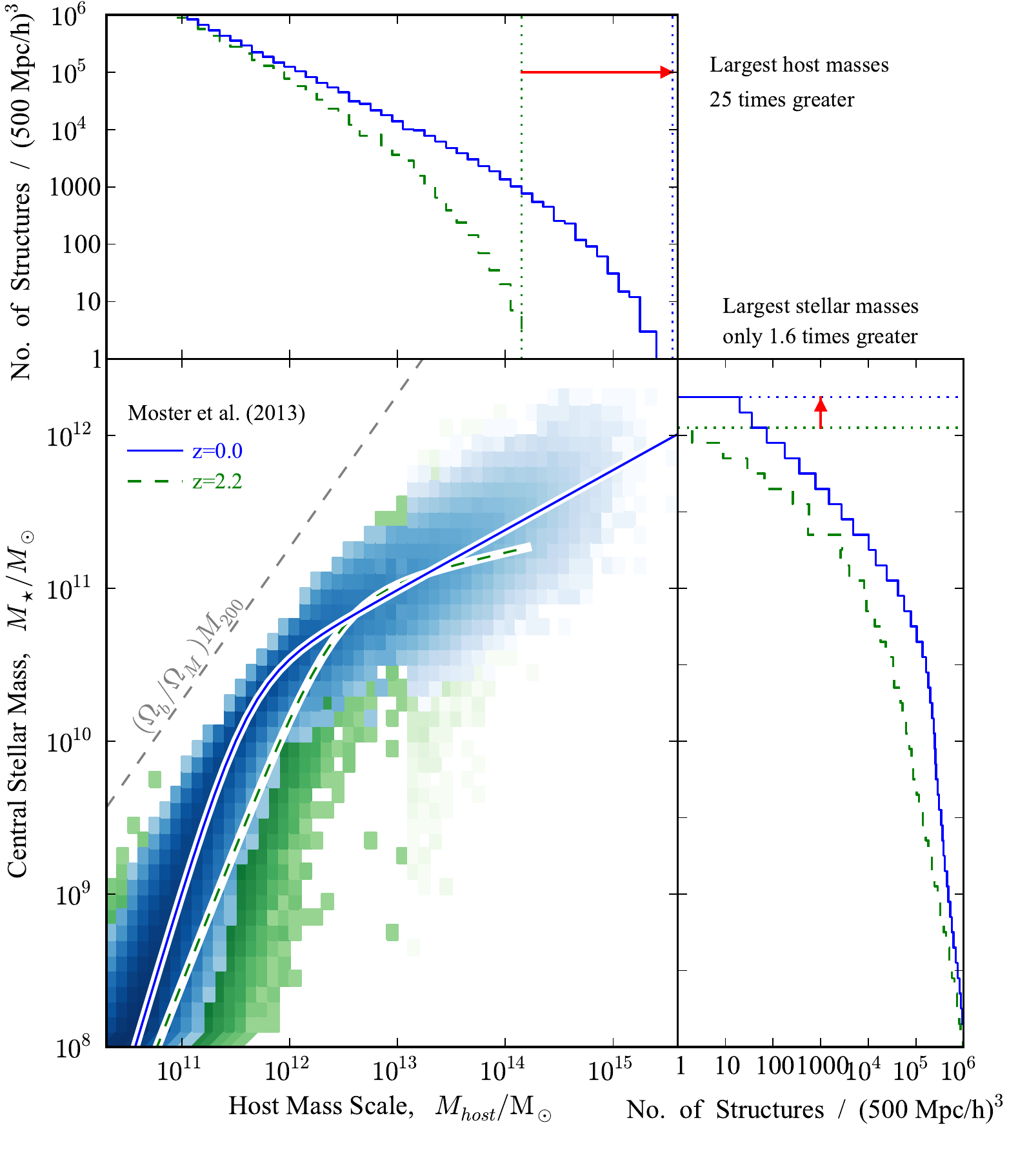}}\label{moster_masses}
\subfigure[The same mapping, but showing host {\em radius}.]{\includegraphics[trim=  212mm 0mm 34mm 18mm, clip, width=0.49\textwidth]{moster_millennium}}\label{moster_radii}
\caption{An illustration of the effects of varying stellar mass content of structures on the $M_\star-R_{200}$ correlation. 
Stellar masses have been assigned to each structure from the simulated volumes (see \S\ref{Simulations}) using the published, empirically motivated occupation function $M_\star(M_{\rm v},z)$ of \protect\cite{Moster13}. The {\bf left panel} shows the resulting distribution in the  $M_\star - M_{\rm v}$ plane at $z=2.2$ (green) and $z=0$ (blue), with the number of structures in the volume projected to the top and to the right. Crucially, at the high mass end, we see that the abundance matching hypothesis implies that galaxies of very similar stellar mass will be hosted by dramatically different structures at the two epochs. The {\bf right panel} translates this to show the host {\em radial} scale as the main x-axis. The decreasing density of structures with time separates the two populations which were overlapped in the left panel, and there is clear evolution of the $M_\star-R_{\rm v}$ trend at all stellar masses. (The $z=2.2$ mock sample {\em does} track the mean, but is very sparse at high-masses due to low halo numbers.)}\label{moster_millennium}
\end{figure*}

The above considerations lead us to believe that cosmology should be able to give us clues on the global structural evolution of galaxies, possibly irrespective of their exact morphology or star formation level. To explore the general consequences of the latter conjecture, we can take the large sample of structures from Fig.\,\ref{millennium} and populate these with galaxies adopting the empirically motivated occupation function of \citet{Moster13}, which provides median stellar mass \citep[based on the initial mass function of][]{Chabrier03} and dispersion for any given host mass. 

Specifically, each structure from the simulation is assigned a stellar mass at random from the distribution given in \citet[][eqn. 2]{Moster13}, drawing a value for each of the four parameters from a normal distribution with the relevant mean and standard deviation \citep[][table 1]{Moster13}. In practice, this amounts to a dispersion in stellar mass of about 0.15 dex.

Using the varying stellar content in these structures allows us to trace the consequences of cosmic evolution through to the galaxy trends.  This connection can also be confronted from the other direction, by taking the observed mass-size relation and showing that this implies a varying galaxy:structure mass ratio \citep{Shen03}. The derived relation can then be used to predict how galaxy sizes evolve with redshift \citep{Somerville08}. The effect of cosmic evolution on the galaxy population has also been investigated by \citet{Firmani09} for the case of both constant $\lambda$ {\em and} a constant average galaxy:structure mass ratio.

The approach taken in this paper was chosen to be complimentary to these existing studies, and to relate our basic {\em analytic} expectation for cosmic structure evolution (\S\S\ref{Introduction}) and stellar mass content (\S\ref{MassContent}) as directly as possible to the galactic radial evolution. Adopting the canonical choice of a constant $\lambda$, is also made to keep the focus on the cosmological origins of any emerging trends. As such, the correlations with host radius derived from these arguments and calculations can be viewed as a 1st order reference point, from which more complex calculations and observations of galactic radii can be interpreted.

Fig.\,\ref{moster_millennium} shows the result of this exercise. The left panels (a) simply represent the inferred relationship between $M_\star - M_{\rm v}$ as published (solid line) with the occupation function, including scatter, applied to simulated structure populations from the two different epochs (shading). The mass functions along each axis are projected to the top and to the right. In this plane, the two populations are overlapping at intermediate masses. But at high mass, the hosts of the most massive galaxies are very different at the respective epochs, as is clear from the mass functions, corroborating the qualitative discussions in \S\ref{Theory} \& \ref{Massive}.

\begin{figure}
\includegraphics[trim=  2mm 100mm 13mm 14mm, clip, width=\columnwidth]{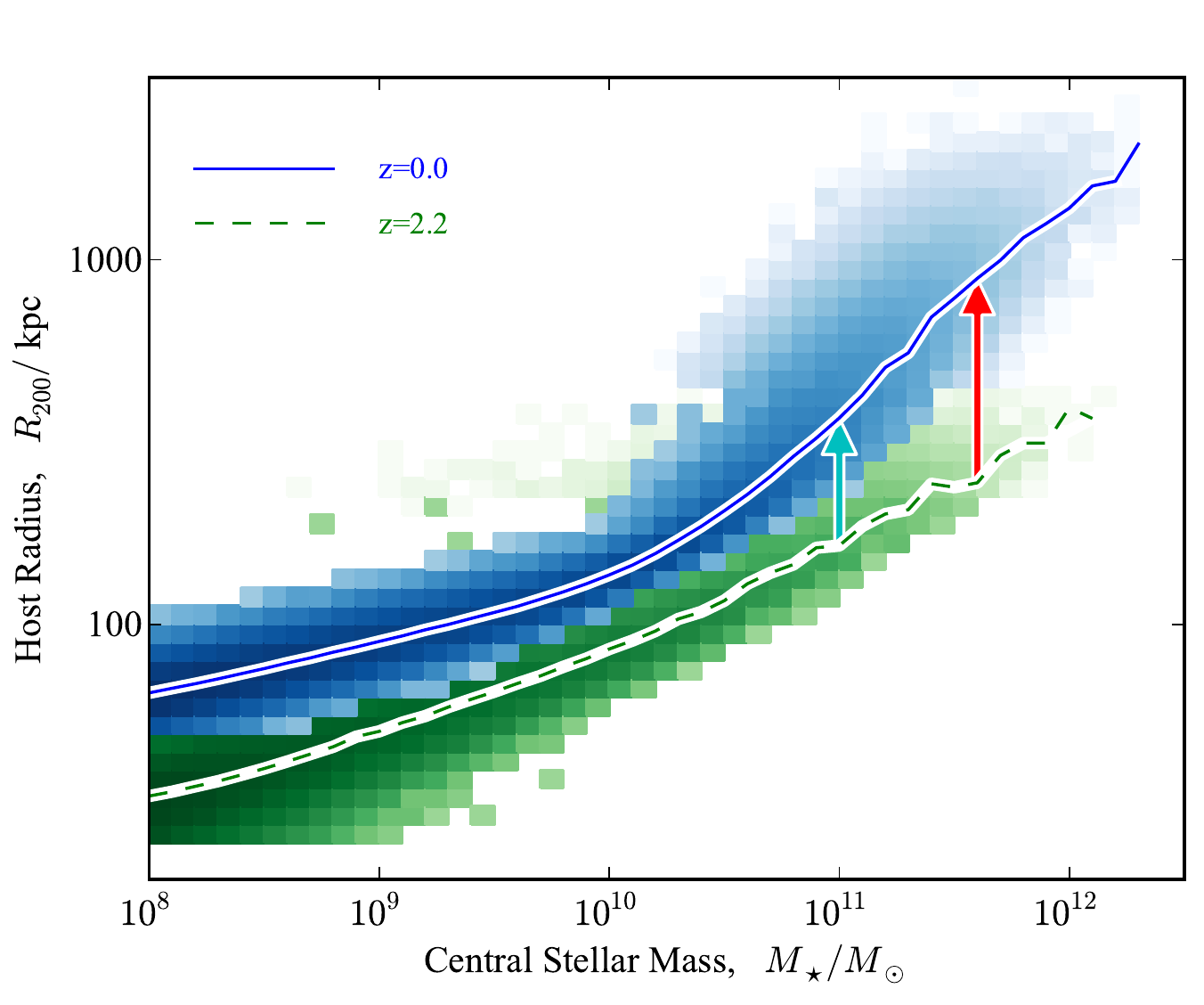}
\caption{The distribution of host radii and central stellar masses for the same mock sample as Fig.\,\ref{moster_millennium}. Shading indicates the number density of all structures at $z=0$ (blue) and $z=2.2$ (green). The thick solid line shows the mean of $\log(R_{\rm 200}/{\rm kpc})$ at each stellar mass interval ($\Delta M_\star=0.1$) and the dashed line shows the same mean at $z=2.2$. Note that this {\em mean radius at a given mass} gives a different correlation to the {\em mean mass at given radius} shown in Fig.\,\ref{moster_millennium}b. The arrows indicate the different evolution of the mean radius at two particular stellar mass intervals, corresponding to the two evolutionary paths shown in Fig.\,\ref{evolution}.}\label{millennium_gals}
\end{figure}

When we translate the x-axis to show host {\em radius}, in the right panels (b), the difference in structure density between the two epochs separates the two populations at all masses, and further accentuates the separation at the high mass end. As indicated in the figure, the number density of structures as a function of their radius evolves dramatically from $z=2.2$ to 0, but the numbers at a given stellar mass evolve much less. So the key assumption behind the mock sample -- that cumulative numbers of both must approximately match -- tells us in this figure that {\em galaxies of the same stellar mass are hosted by structures with dramatically different radii at different epochs}.

This effect is even more pronounced than Fig.\,\ref{moster_millennium}b might indicate. To appreciate this, it is important to consider the difference between a `mean stellar mass for a given host' and a `typical host for a given stellar mass'. For the rapidly declining number densities of structures at the high mass end, this is particularly important.

This difference is illustrated by comparing Fig.\,\ref{moster_millennium}b with Fig.\,\ref{millennium_gals}, which shows exactly the same mock sample but with the axes reversed so that stellar mass runs along the x-axis, as is usually the choice for plotting observational samples. Crucially, the mean line which is plotted is now in bins of $M_\star$ (not $R_{200}$). 

\begin{figure}
\includegraphics[trim=  2mm 100mm 13mm 14mm, clip, width=\columnwidth]{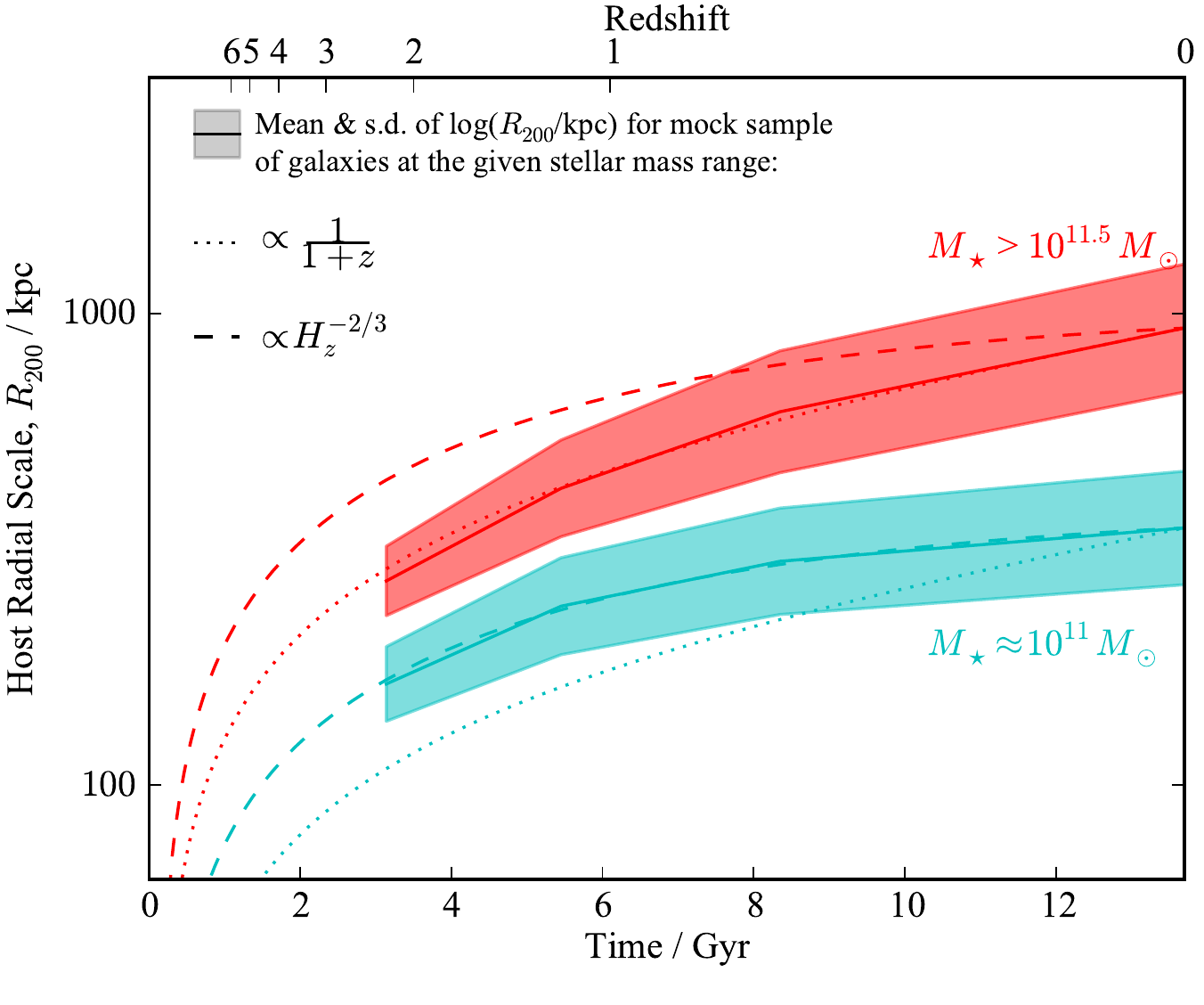}
\caption{This figure takes galaxies in two mass ranges from the same mock sample as Figs. \ref{moster_millennium} \& \ref{millennium_gals}, and shows the radii of the structures which host each category as a function of time. As can be seen from Fig.\,\ref{millennium_gals}(a), $M_\star\approx10^{11}M_\odot$ galaxies are always hosted by structures of the same mass. This is why the mean radii of this category tracks the cosmological expansion, $\propto H_z^{-\nicefrac{2}{3}}$ (see eqn. \ref{RvMv}). The hosts of the most massive galaxies are more massive at later times, which is why the radii of the $M_\star>10^{11.5}$ category increases more steeply, following $1/(1+z)$.}\label{evolution}
\end{figure}

This shows that host radii for a given stellar mass are evolving across the entire range, and {\em particularly} at high mass (though the opposite might have been concluded by glancing at the mean lines in Fig.\,\ref{moster_millennium}b). This evolution is indicated in the figure for two particular stellar mass ranges, for which more details are plotted alongside in Fig.\,\ref{evolution}. This compares the mean and standard deviation of the inferred hosts of $10^{11}$ and $10^{12}\Msun$ galaxies as a function of time. For comparison, dashed lines show the evolution of the radius of host structures of the same {\em total} mass, derived in eqn. \ref{Dv}. 

An appreciation of these changing host populations, for galaxies of a given mass, is certainly of independent value. But this discussion is of course leading back towards a consideration of galactic radii. {\em If} it is indeed the case that galactic radii scale with their host structure, then it is possible to interpret Fig.\,\ref{millennium_gals} as an explanation of the trend in radius as a function of stellar mass and its evolution with  time, and Fig.\,\ref{evolution} as showing that this leads to the increase with time of the mean radius of galaxies of a given stellar mass.

\begin{figure}
\includegraphics[trim=  2mm 12mm 13mm 135mm, clip, width=\columnwidth]{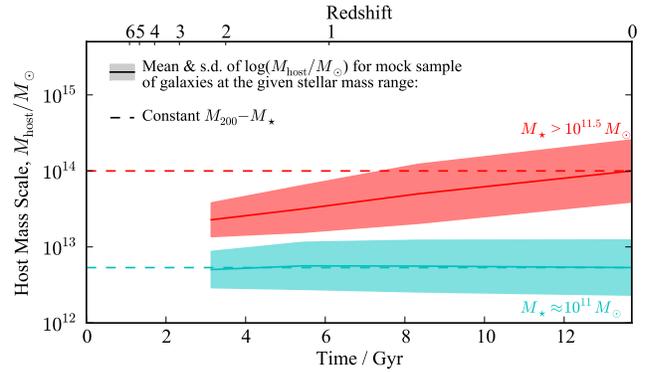}
\caption{Confirmation of the arguments presented in Fig.\,\ref{evolution} concerning the mass of structures which host galaxies of the given two mass ranges at different epochs. The abundance-matching hypothesis which generates the mock sample implies that galaxies with $M_\star\approx10^{11}M_\odot$ are always hosted by structures of $M_{200}\sim5\times 10^{12}M_\odot$, whereas the most massive galaxies, with only weak correlation to their host, are statistically more likely to be found in more massive hosts as time goes on.}\label{evolution_mass}
\end{figure}

As shown in Fig \ref{evolution_mass}, the abundance-matching hypothesis indicates that $M_\star\approx 10^{11}\Msun$ galaxies are found in similar mass hosts at all epochs. Because of this, the evolution in their host radii almost exactly tracks the cosmic evolution, $\propto H_z^{-\nicefrac{2}{3}}$. The most massive galaxies, however, are hosted by more massive structures as the universe evolves, so the increase in the mean host radii of galaxies with $M_\star>10^{11.5}\Msun$ {\em exceeds} the background cosmic expansion.


In conclusion of \S\ref{Mapping}, this is a good juncture to re-enforce once more the distinction between understanding the growth of individual objects and understanding changes in the whole population with time. This was addressed at the end of \S\ref{Individuals} for cosmic structures themselves, and we can now redress this question in terms of their expected stellar mass content. The occupation function used here implies that about $440$ structures in the simulation volume would host galaxies exceeding $10^{11.5}\Msun$ at $z=2.2$. But at $z=0$, over 4100 structures would be expected to host galaxies in this category. So this analysis suggests that, in the same volume, a low redshift sample at $M_\star>10^{11.5}\Msun$ is almost entirely ($\sim90$ per cent) composed of {\em new arrivals in this category} since $z\sim2$. Of course, in the observational scenario, the larger survey volumes at high redshift will go some way to alleviate this. Nonetheless, a full appreciation of this effect is surely essential for any observational analysis that wishes to associate high redshift progenitors with local descendants, or understand the variation of mean values within fixed stellar mass ranges.

To conclude \S\ref{Galaxies} as a whole, the theoretical picture that we are motivating with the arguments and results above is this: Whilst the {\em mass} extant at the centre of a structure can be affected by feedback and cooling, thus warping -- or even losing -- the correlation between with the host and stellar mass, the {\em specific angular momentum} of the central material remains indelible, and thus tied to that of structure from which it cooled. This would  be clearly true in the simple case where feedback were just {\em indiscriminate} in removing material. And in the general case the argument is -- if anything --  strengthened, given that feedback will preferentially eject {\em low} angular momentum material, leaving the radius and velocity of the outermost cooled gas \& stars with a strong residual correlation to the host.

Whilst the total stellar mass of galaxies is controlled by feedback (in the intermediate, efficient cooling regime) and cooling limits (at limiting high and low masses), the stellar orbits still carry the imprint of the structures intrinsic specific angular momentum. Thus, it may be possible to view the observed trends in galactic radii as a {\em direct correlation in radius} to their host structures alongside a {\em varying stellar mass content}.

\section{Observational context}\label{Application}

\begin{figure*}
\includegraphics[trim=  24mm 2mm 29mm 3mm, clip, width=\textwidth]{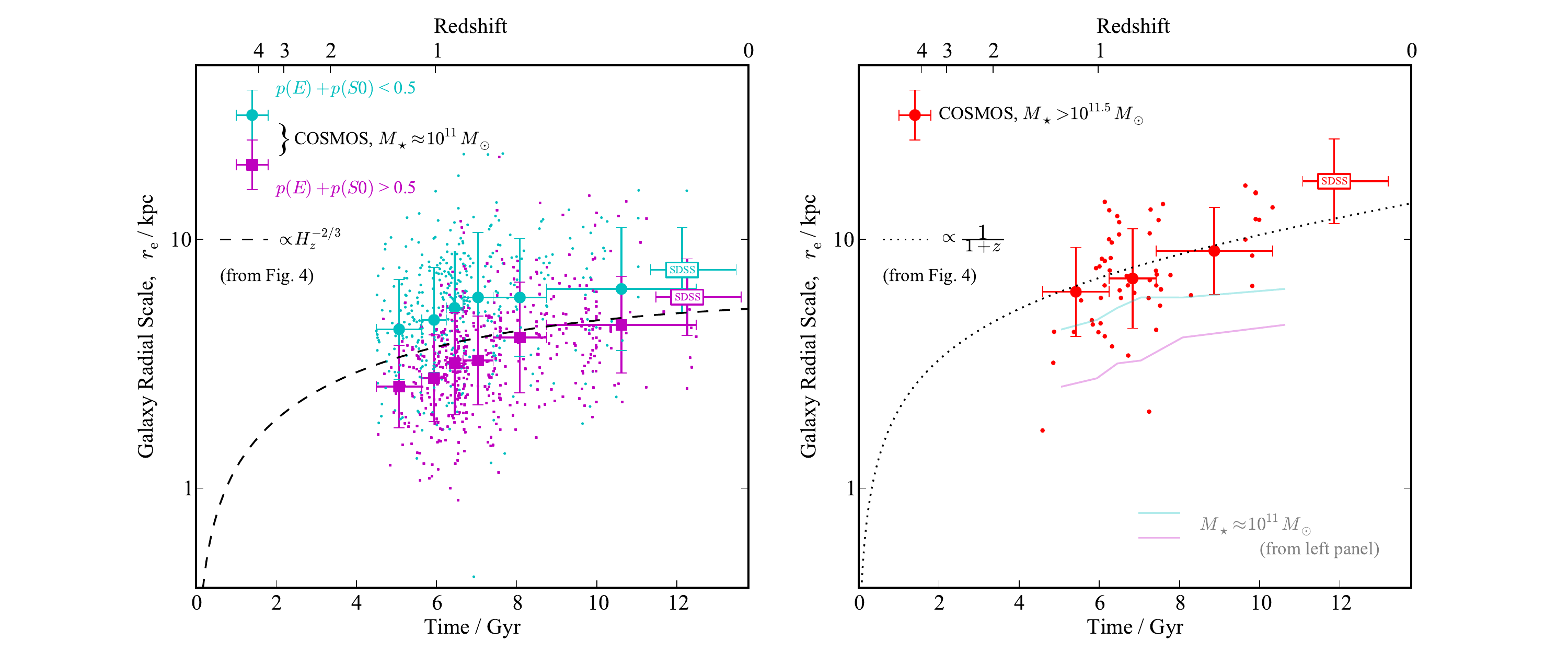}
\caption{Observational size estimates for galaxies in the COSMOS and SDSS surveys which fall into two stellar mass ranges, shown as a function of time according to their photometric redshift. Those with $z>1.5$ are deemed unreliable and are not shown here. The {\bf left panel} shows 858 galaxies with estimated stellar masses in the range of $10^{10.8}\le M_\star/\Msun<10^{11.2}$ (small points). This sample is then divided into 6 groups in time (with 143 galaxies in each group) and then further divided according to the galaxies assigned probability of belonging to the E or S0 morphological category. Errorbars show the mean and standard deviation of radii in each subsample. The {\bf right panel} shows 57 galaxies with estimated stellar mass $M_\star \ge 10^{11.5}\Msun$ (the highest being $7 \times 10^{11}\Msun$ and the mean $4\times 10^{11}\Msun$). The dashed lines in both panels indicate the predictions from Fig.\,\ref{evolution} for the relevant mass range, following the theoretical arguments of \S\ref{Structures}--\ref{Galaxies} and analysis of mock samples. This indicates that the observed evolution is consistent with the theoretical ideas discussed in this work. The additional result from this figure is that, whilst the different morphological types do have slightly different mean radii at any given epoch, the relative {\em evolution} of these means does not differ significantly (see also Fig.\,\ref{evolution_zmorph}). }\label{evolution_obs}
\end{figure*}

To confront the main theoretical predictions of \S\ref{Structures} and \S\ref{Galaxies} with observations, we present in Fig.\,\ref{evolution_obs} the evolution of effective radii of sub-sample of galaxies from the COSMOS survey \citep{Huertas13}. The two panels show the sizes of galaxies at two different stellar mass ranges ($10.8<\log (M_\star/M_\odot)<11.2$ and  $\log (M_\star/M_\odot)>11.5$) as a function of time, based on their photometrically determined redshifts. These stellar mass ranges are chosen to be as large as possible without the mean mass in the range varying significantly as a function of time (the mean mass in both ranges varies by less than 0.03 dex). For the low mass range, we also split the sample based on their probable morphology, along $P(E)+P(S0)=0.5$. We do not make this split in the high mass range, where only 11 of the 57 in this much smaller sample have $P(E)+P(S0)<0.5$ (and  none less than 0.1).

As explained by \citet{Huertas13}, the sample is made of group \citep[from the group catalog of][]{George11} and field galaxies and is complete in stellar mass down to $M_\star\approx10^{10.5}\Msun$ at $z\sim1$ (notice that not only central galaxies are included in this sample). Galaxy sizes have been computed with {\sc galapagos} \citep{Barden12} and stellar masses are derived through SED fitting with \citet{Bruzual03} synthesis population models, using all the COSMOS filters and assuming a Chabrier IMF \citep{Bundy06}. Finally, morphologies have been computed with {\sc galsvm} \citep{Huertas08} and extensively checked \citep{Huertas13}.

We also show in Figs.\,\ref{evolution_obs} and \ref{evolution_z} the values at $z\sim0$ from the Sloan Digital Sky Survey DR7. Sizes are computed by fitting a 2D single Sersic profile as for the COSMOS sample \citep[see][for more details]{Meert13} and morphologies are taken from the morphological catalog of \citet{Huertas11} using the same automated algorithm, {\sc galsvm}, used for the high redshift sample. Mass to light ratios have been obtained from the MPA-JHU DR7 release. They are derived through SED fitting using BC03 synthesis population models \citep{Bruzual03} and a Kroupa IMF following the procedure presented in \citet{Kauffmann03} and \citet{Salim07}. We then convert to stellar masses by multiplying the M/L of each galaxy by its luminosity estimated from the best fit Sersic model and convert to a Chabrier IMF in order to be consistent with the high redshift sample.

Included with the observational points and means in Fig.\,\ref{evolution_obs} are the same two theoretically-motivated lines that were plotted in Fig.\,\ref{evolution} with the mock galaxy sample, adjusted to apply to galactic radii using $R_{\rm gal}=\lambda R_{\rm v}$ with $\lambda\approx 0.015$ (\S\ref{rgal_rv}). If Fig.\,\ref{evolution_obs} were intended for a rigourous assessment of a precise prediction, it would be necessary to incorporate a scatter in $\lambda$ and statistically assess both the mean and scatter predicted by the theory against observations. This, we leave for subsequent studies.

At the level of proposing an accessible physical explanation of the basic trend, which is the goal here, the agreement between the basic theoretical prediction and the observational estimates, in both panels of Fig.\,\ref{evolution_obs}, is consistent with the idea that they are mirroring the cosmological expansion via their hosts. The change in mean size of the $M_\star\approx10^{11}\Msun$ galaxies, in particular, tracks the change in the equivalent host radii of the mock sample very closely, as further confirmed in Fig.\,\ref{evolution_z}. 

At the highest stellar mass range, the change in size with time is indeed steeper than at $M_\star\approx10^{11}\Msun$. The slightly lower mass sample are found in halos of the same mass at all redshifts, and these halos are progressively larger -- with more specific angular momentum -- as time goes on, following (\ref{RvMv}). The hosts of the highest mass sample are also larger for this same reason  but, {\em additionally}, they tend to form in more massive structures at later times (Fig.\,\ref{evolution_mass}). 

To summarise, there are two effects causing the highest mass category to be larger with time: 
\begin{itemize}
\item{Host structures at a given mass getting larger, tracking the cosmic expansion (eq. \ref{RvMv}) and}   
\item{the most  massive galaxies being likely to form in progressively more massive structures as larger regions of the Universe collapse (Fig.\,\ref{evolution_mass}).}
\end{itemize}

Fig.\,\ref{evolution_obs} shows that the results of the survey are consistent with the prediction from mock sampling, though looking at the sample as a function of $(1+z)$ in Fig.\,\ref{evolution_z} gives some indication that the evolution may be somewhat steeper still, a simple regression fit favouring $\gamma \approx 1.8$. But a larger sample is required to conclude firmly.

The dependence of the size evolution on stellar mass has also been discussed in several previous observational works with different results. \citet{Williams10} and \citet{Ryan12} measured a mass dependence where the radii of massive galaxies are found to change more with redshift, leading to a changing slope in the mass-size relation. Other work such as \citet{Damjanov11} or \cite{Newman12} suggests that the slope of the mass-size relation is mass-independent. \citet{Huertas13} showed that the correlation with mass that is concluded is dependent on how the selection is performed. In fact, the authors showed that when pure bulges are selected, the dependence on mass seems to be more pronounced than for all passive ellipticals. 

Recent observational papers have also studied the impact of environment on the size evolution. While it is still uncertain, several works have reported that sizes of ETGs are larger in dense environments at $z>1$ \citep[e.g][]{Delaye13,Lani13,Papovich12}, which does not seem to be the case at $z\sim0$ \citep{Poggianti13,Huertas13}. We do not try confront these early results in this particular study, but do note that the approach taken here, using the simulated structure population as a mock sample, could be easily extended to investigate different environments, something we hope to pursue in future work.

Concerning the sizes for different morphologies, Fig.\,\ref{evolution_z} shows that the most disk-like half of the sample are, on average, double the size of their spheroidal counterparts. But the samples are overlapping and the respective means are still within a standard deviation of each other. The change of the mean radius in time is visibly very similar for both subsamples. A simple regression analysis does yield a slightly different dependence on $(1+z)$, suggesting $\gamma=0.9$ and 1.2 for disk-like and spheroidal samples respectively (Fig.\,\ref{evolution_zmorph}). However, given the scatter, we do not consider this difference significant. Similar results are found when dividing the sample by Sersic index or star formation rate tracer (Fig.\,\ref{evolution_alt}). So the galaxy size evolution, in these terms at least, appears to be similar for different categories, consistent with the theoretical arguments presented in \S\ref{Structures} and \S\ref{Galaxies}.

\citet{Newman12} also found that star forming and passive galaxies ($M_\star>10^{10.5}\Msun$) evolve in a similar way \citep[see also][]{Law12}. However, \citet{Buitrago08} found that galaxies with $M_\star>10^{11}M_\odot$ and Sersic index, $n_{\rm s}<2.5$ evolve less than galaxies with similar mass but higher Sersic index. This is not confirmed with the sample used in this work. \citet{Barden05} and \citet{vanDokkum13} also reported a steeper degree of evolution for ellipticals than for disks. The differences between these works might come from the different selections (mass bins, morphologies, number density, etc.) but certainly requires further investigation which is beyond the scope of the present theoretical paper.

\section{Summary}\label{Summary}

In this article, we began by reviewing the sizes of collapsed structures predicted by standard cosmology, how this will lead to an instantaneous trend in the masses and radii of this population, and that this trend evolves with time (\S\ref{Structures}). We emphasised also the important distinction between the evolution of this trend and the evolution of any individual structure (Fig.\,\ref{millennium}). 

In \S\ref{rgal_rv}, we reviewed recent work by \citet{Kravtsov13} and \citet{Kassin12} which supports the idea that galactic radii will be directly correlated to their hosts, and the established theoretical reasoning that this might be the case \citep{Mo98}. Conversely, it is equally well established empirically that galactic {\em masses} do {\em not} correlate directly with those of their hosts but vary greatly depending on the mass range in question. The physical causes of this variation were reviewed in theoretical discussions in \S\ref{MassContent}. We then show how these physical limits on stellar mass content (Fig. \ref{sketch}) will translate to limits in host radii (Fig. \ref{sketch_radius}) and, emphasising the effect of greater structures collapsing at progressively later times, to expected limits on galactic radii that are broadly consistent with half-light radii found in the SDSS and local dwarf spheroidals (Fig. \ref{sketch_data}).

To further quantify this, and better model the highest mass population for which scatter and sampling become particularly important, the sample of structures from the Millennium simulations shown in \S\ref{Structures} \citep{Lemson06} were populated in \S\ref{Mapping} with galaxies according to the empirically-motivated occupation function of \citet{Moster13}. This exercise illustrates (Fig.\,\ref{moster_millennium}) how even a relatively static distribution in stellar mass--host mass (emergent at $M_\star\approx 10^{11}\Msun$) will still translate to a stellar mass--{\em radius} correlation which evolves strongly with time ($\propto H^{-\nicefrac{2}{3}}$) due to cosmological evolution reviewed in \S\ref{Theory}. 

The highest stellar mass category ($\gtsim 3\times10^{11}\Msun$), being hosted by steadily more massive structures as the universe evolves, will therefore show further accentuation of this trend (Fig.\,\ref{evolution}). Also, we find that this extension of the abundance matching hypothesis implies that a striking $\sim90\%$ of the hosts of massive galaxies at $z\approx0$, in this constant comoving volume, joined the category since $z\approx2.2$.

\begin{figure}
\includegraphics[trim=  5mm 2mm 15mm 3mm, clip, width=\columnwidth]{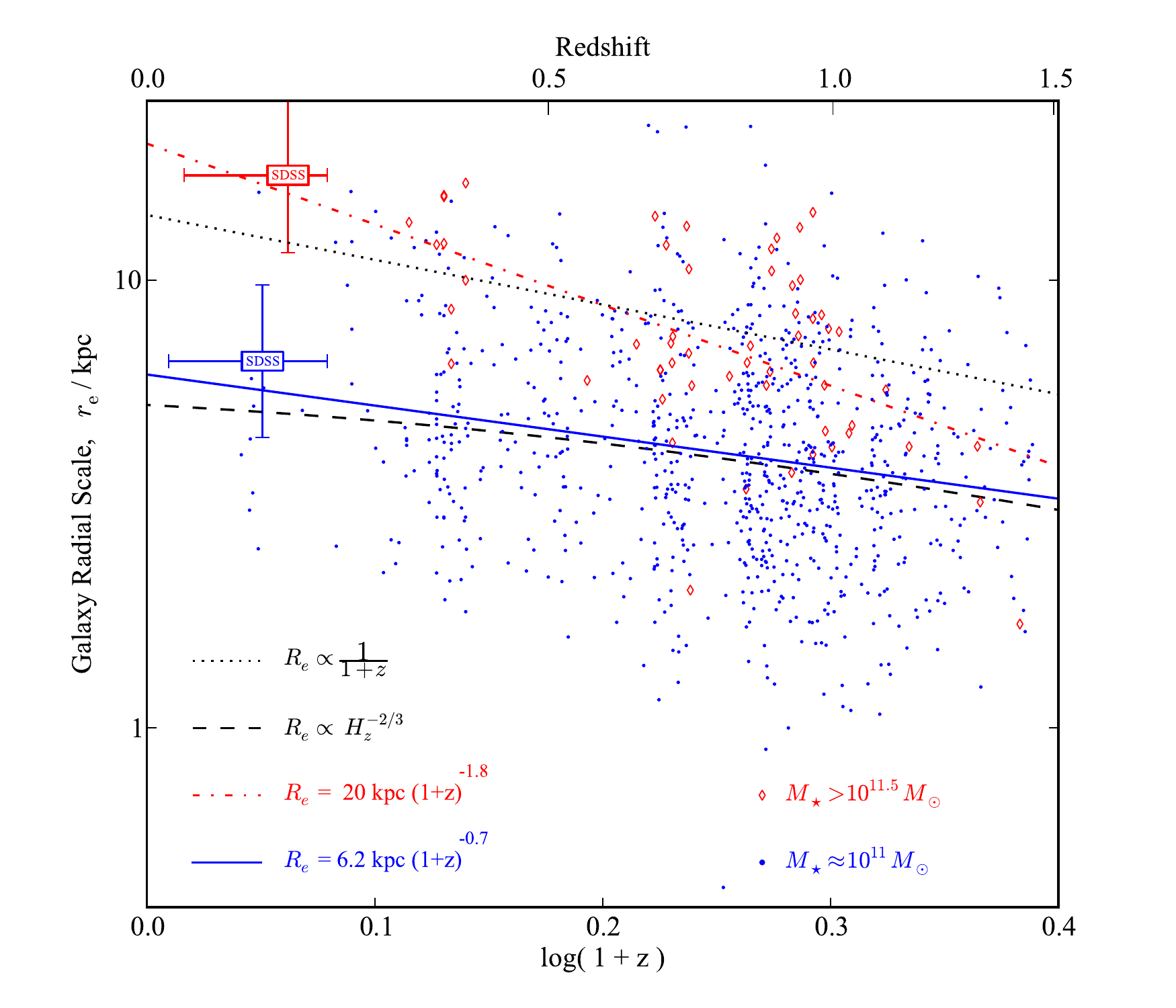}
\caption{The estimated radii and redshifts for galaxies in the COSMOS sample that were presented in Fig.\,\ref{evolution_obs}, but now plotted as a function of $\log(1+z)$, and showing the results of simple 1st-order regression fit of the data on these axis. The solid and dot-dashed lines are fits to the COSMOS data alone (not the SDSS), for the two subsamples in stellar mass range as shown in the key. The dashed and dotted lines show the expected correlation from the theoretical arguments presented in \S\ref{Theory} and \S\ref{Galaxies}.}\label{evolution_z}
\end{figure}

In \S\ref{Application} we investigated the applicability of this idea to galaxies from the COSMOS and SDSS surveys, finding the evolution in radius of the stellar mass ranges that have been observed is consistent with the expected theoretical evolution summarised above, without marked dependance on morphological type (Figs.\,\ref{evolution_obs}\,\&\,\ref{evolution_z}). This enforces the key point of the article; that {\em if} the correlation between host and galactic radius is indeed robust, then our existing understanding of the varying galactic {\em stellar mass} content of structures extends automatically to explain evolving trends in galactic {\em radii}. 

Thus, we conclude that it may be possible to understand these trends in galactic radii, and their evolution, as arising from: 

\begin{itemize}
\item{Collapsed structures carrying an imprint of the cosmology,}
\item{with stellar mass content that varies strongly with host mass,}
\item{but mean scalar specific angular momentum that remains directly correlated.}
\end{itemize}

\section*{Acknowledgements}
This research was funded by ERC grant number 267399, {\em Momentum}, and FS acknowledges support from a Marie Curie grant. The Millennium Simulation databases used in this paper and the web application providing online access to them were constructed as part of the activities of the German Astrophysical Virtual Observatory (GAVO), and we are grateful to Gerard Lemson for his assistance with these databases. The galaxy sizes from the SDSS were used with the kind permission of Mariangela Bernardi and Alan Meert, and the COSMOS data with that of Simona Mei. Thanks go to all of them, together with Ignacio Trujillo, for their many helpful comments on the manuscript. Finally, MJS would like to thank Andrew Benson for several constructive discussions during the course of the project.

\bibliographystyle{mn2e_Daly}
\bibliography{references}

\appendix

\section{Additional observational details}\label{Details}

\begin{figure}
\includegraphics[trim=  5mm 2mm 15mm 3mm, clip, width=\columnwidth]{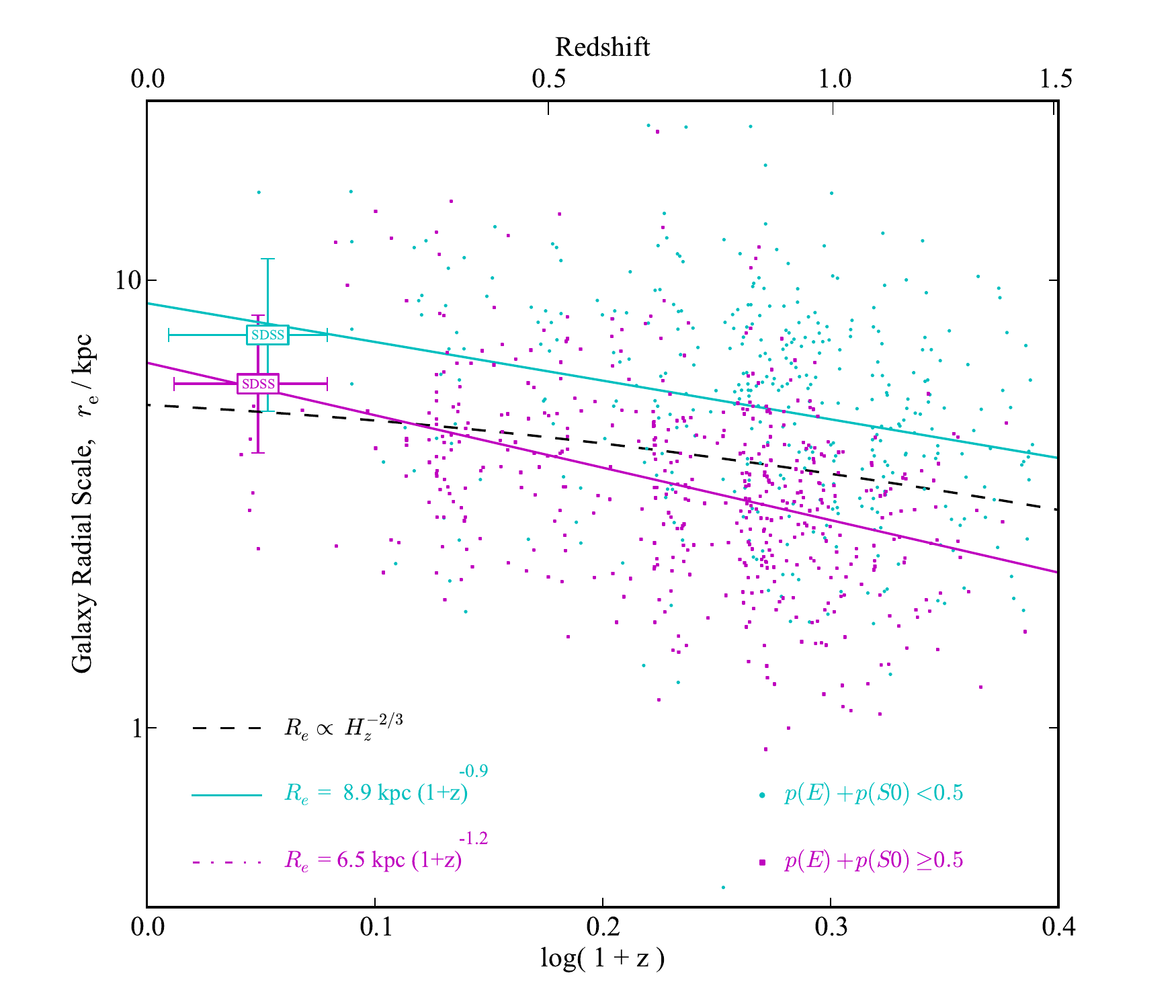}
\caption{The same observational sample as Fig.\,\protect\ref{evolution_z}, but with the two morphological categories separated. The linear fits to $\log(1+z)$ are now calculated for each subsample separately (without additional constraint from the SDSS results, shown only for reference). Though straightforward linear regression does find some difference in the redshift dependence of sizes in each category, the subsamples completely overlap each other at all redshifts.}\label{evolution_zmorph}
\end{figure}

For completeness, this section shows some additional analysis of the COSMOS data that was represented in the main article in Figs. \ref{evolution_obs} and \ref{evolution_z}. The first of these, Fig.\,\ref{evolution_zmorph}, shows the $M_\star\approx 10^{11}\Msun$ sample divided into the same two categories as Fig.\,\ref{evolution_obs}, but plotted on the axes of Fig.\,\ref{evolution_z}. This includes a basic regression fit which shows similar, if slightly differing evolution for the two sub-samples.

\begin{figure}
\subfigure{\includegraphics[trim=  210mm 73mm 20mm 8mm, clip, height=0.36\columnwidth]{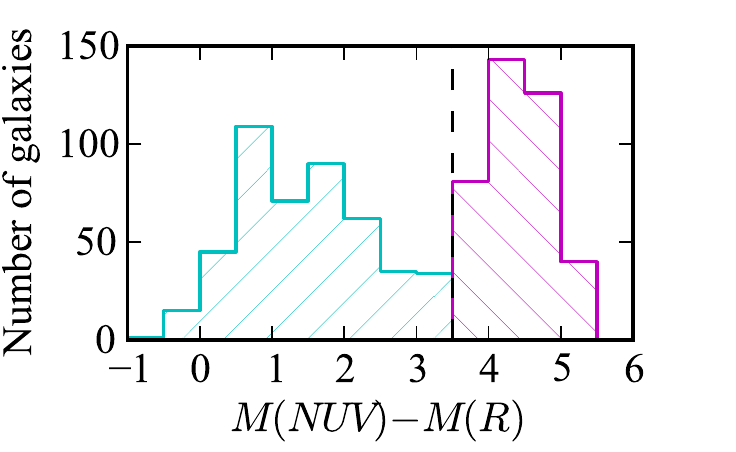}}
\subfigure{\includegraphics[trim=  210mm 73mm 28mm 8mm, clip, height=0.36\columnwidth]{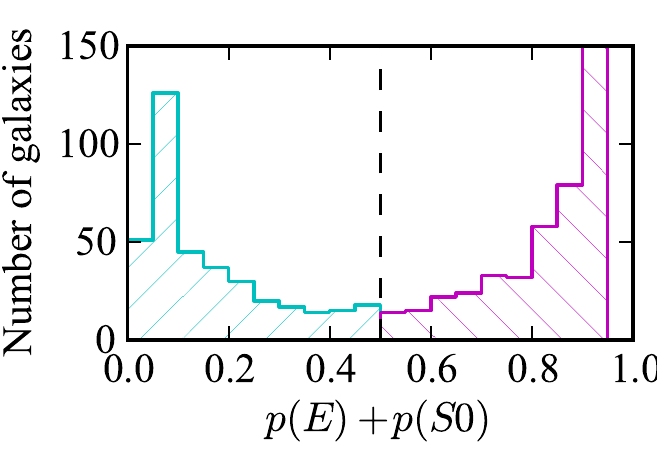}}
\subfigure{\includegraphics[trim=  222mm 73mm 30mm 8mm, clip, height=0.36\columnwidth]{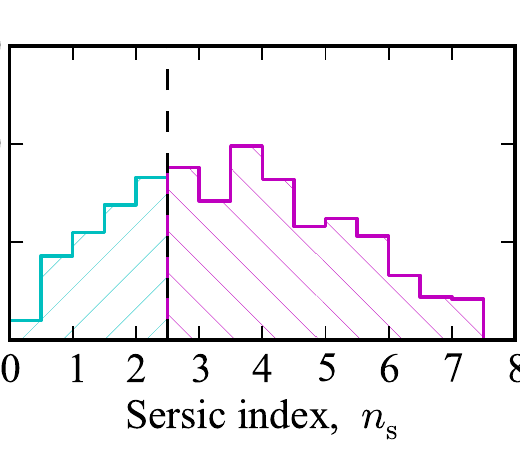}}
\caption{The distribution of galaxies as a function of the three alternative criteria that have been shown in Figs. \ref{evolution} and \ref{evolution_alt}. Dividing by star formation rate tracer (upper left panel) indicates sufficient bi-modality to reassure us that the division into sub-populations along $M(UV)-M(R)=3.5$ is appropriate. However, the distribution by Sersic index (bottom right panel) indicates no compelling bi-modality. Division by probable morphology (bottom left panel), as used throughout the article, is highly bi-modal and thus the least sensitive of the three to the choice of dividing criteria.}\label{distributions}
\end{figure}

Fig.\,\ref{evolution_alt} shows two alternative versions of Fig.\,\ref{evolution_obs}, dividing by Sersic index or star formation tracer instead of morphology. The distributions of the sample as a function of each of these possible criteria is also shown for reference in Fig.\,\ref{distributions}.

\begin{figure*}
\subfigure{\includegraphics[trim=  24mm 2mm 154mm 3mm, clip, width=0.49\textwidth]{evolution_sfr}}
\subfigure{\includegraphics[trim=  20mm 2mm 158mm 3mm, clip, width=0.49\textwidth]{evolution_ns}}
\caption{Alternative versions of Fig.\,\ref{evolution_obs}, using the same format but dividing the sample according to a different criterion. The {\bf top panel} divides the sample according to the star formation rate tracer, $M(UV)-M(R)$, the {\bf bottom panel} divides it  according to the Sersic index. }\label{evolution_alt}
\end{figure*}

\label{lastpage}

\end{document}